\documentclass[12pt]{iopart}
\usepackage{iopams}
\usepackage{graphicx}

\expandafter\let\csname equation*\endcsname\relax 
\expandafter\let\csname endequation*\endcsname\relax 
\usepackage{amsfonts}
\usepackage{amsmath}
\usepackage{bm}

\usepackage{cite}

\newcommand{\<}{\langle}
\renewcommand{\>}{\rangle}
\newcommand{\apj}{Astrophys. J.}

\newcommand{\mnras}{Mon. Not. Roy. Astron. Soc.}

\newcommand{\CD}{{\cal D}}
\newcommand{\be}{\begin{equation}}
\newcommand{\ee}{\end{equation}}
\newcommand{\bea}{\begin{eqnarray}}
\newcommand{\eea}{\end{eqnarray}}
\newcommand{\sfrac}[2]{{\textstyle{#1\over#2}}}

\newcommand{\average}[1]{\left\langle #1 \right\rangle_{\CD}}
\newcommand{\saverage}[1]{\langle #1\rangle}
\def\la {\langle}
\def\ra {\rangle}

\renewcommand{\div}{\mbox{div}\,}

\def \D {\mathrm{D}}
\def \d {\mbox{d}}

\def\c {\mbox{curl}\,}
\def\ep {\varepsilon}

\def\a {a}
\def\b {b}

\newcommand{\obinna}[1]{\textsc{\Large #1}}
\renewcommand{\obinna}[1]{}

\begin{document}

\title{Is backreaction really small within concordance cosmology?}
\author{Chris Clarkson and Obinna Umeh}

\address{Astrophysics, Cosmology \& Gravity Centre, and, Department of Mathematics and
Applied Mathematics, University of Cape Town, Rondebosch 7701, Cape Town, South Africa}

\ead{\mailto{chris.clarkson@uct.ac.za}, \mailto{obinna.umeh@uct.ac.za}}

\begin{abstract}

Smoothing over structures in general relativity leads to a renormalisation of the background, and potentially many other effects which are poorly understood. 
Observables such as the distance-redshift relation when averaged on the sky do not necessarily yield the same smooth model which arises when performing spatial averages.
These issues are thought to be of technical interest only in the standard model of cosmology, giving only tiny corrections.
However, when we try to calculate observable quantities such as the all-sky average of the distance-redshift relation, we find that perturbation theory delivers divergent answers in the UV and corrections to the background of order unity. 
There are further problems. Second-order perturbations are the same size as first-order, and fourth-order at least the same as second, and possibly much larger, owing to the divergences. Much hinges on a coincidental balance of 2 numbers: the primordial power, and the ratio between the comoving Hubble scales at matter-radiation equality and today.
Consequently, it is far from obvious that backreaction is irrelevant even in the concordance model, however natural it intuitively seems.

\end{abstract}

\maketitle



\section{Introduction}

The standard Friedmann-Lema\^{\i}tre-Robertson-Walker (FLRW)
models of cosmology can account for
all observations to date with just a handful of parameters~-- a resounding success. These
are the simplest reasonably realistic universe models possible
within general relativity: homogeneous, isotropic, and flat to a
first approximation, with a scale-invariant spectrum of Gaussian
perturbations from inflation added on top to account for structures down to the
scale of clusters of galaxies. 
Were there a physical motivation for the
value of the cosmological constant and the associated coincidence
problem we would have a nearly complete understanding at nearly all epochs. 

The coincidence problem has led some to speculate that the evidence for a cosmological constant might in fact be a misunderstanding of the standard model, and that complex relativistic `backreaction' effects are present when structure forms. This might be important in a variety of ways~\cite{Ell84,Buchert:1995fz,Buchert:1999er,Buchert:2001sa,Rasanen:2003fy,Barausse:2005nf,Kolb:2005da,Rasanen:2006kp,Kasai:2007fn}. The canonical mechanism might arise by altering the dynamics of spacetime when we smooth over structure to form a spatially homogeneous and isotropic `background'~\cite{Russ:1996km,Kolb:2004am,Li:2007ci,Li:2007ny,Li:2008yj,Clarkson:2009hr,Clarkson:2009jq,Chung:2010xx,Umeh:2010pr}. In the standard picture, this would be represented by a mismatch between the background model at early versus late times~-- in the standard model they are assumed to be the same.  Alternatively, smoothing observations over the sky might lead to a significantly different cosmology than a corresponding spatial smoothing process (sometimes known as the fitting problem~\cite{EllSto87}). Some authors have argued that it is specifically an infra-red divergence which appears at second-order that gives the possibility of the backreaction effect which can mimic dark energy~\cite{Kolb:2005da,Kolb:2004am,Barausse:2005nf}. 

That such mechanisms might lead to significant effects~-- let alone be the cause of apparent acceleration~-- has been quickly dismissed by many in the community~ \cite{Ishibashi:2005sj,Flanagan:2005dk,Hirata:2005ei,Geshnizjani:2005ce,Bonvin:2005ps,Siegel:2005xu,Giovannini:2005sy,Bonvin:2006en,Vanderveld:2007cq,Behrend:2007mf,Kumar:2008uk,Rosenthal:2008ic,Krasinski:2009qq,Ziaeepour:2009zu,Tomita:2009ar,Baumann:2010tm,Green:2010qy}. We can understand the argument why backreaction must be small as follows. 

The fields that are propagating on the background must alter its dynamics through the non-linearity of the field equations. According to observers at rest with the gravitational field, the potential itself is small everywhere outside objects less dense than neutron stars, and so we write $g_{ab}=g^{(0)}_{ab}+g^{(1)}_{ab}$ where $g^{(1)}_{ij}\sim\Phi g^{(0)}_{ij}$ where $\Phi\sim10^{-5}\ll1$. The affine connection or Ricci rotation coefficients determine the dynamics of the spacetime, and are generically $\mathcal{O}(\partial \Phi)$; the field equations are $\mathcal{O}(\partial^2 \Phi)$. A perturbation of wavelength $\lambda= a/k$, where $k$ is the comoving wavenumber and $a$ is the scale factor, much less than the Hubble scale can give rise to large fluctuations in the field equations $\mathcal{O}(k^2 \Phi)$, even though the change from the background metric is small. Such terms describe density fluctuations, and  these can be large even though the metric potentials are small. After averaging over a large domain, however, all first-order quantities vanish by assumed homogeneity and isotropy of the perturbations. This picture predicts the CMB temperature anisotropies and the matter power spectrum to all required precision.  At second-order the Einstein tensor can only contain terms as large as the peculiar velocity squared $v^2\sim(\partial\Phi)^2$. The amplitude of these terms are only the same size as $\Phi$, with contributions from higher derivative terms being contained in pure divergence terms which are zero by statistical homogeneity. Higher-order terms are suppressed by further factors of $\Phi$. Any infra-red contribution must be a gauge effect, and can be renomalized away.  And so, backreaction cannot be significant. 

Much of this argument is correct. Yet it also contains elements which are startling: \emph{second-order terms are the same size as first-order}. In particular, the second-order potential has amplitude $\Phi^{(2)}\sim\Phi$, which is really not supposed to happen! Furthermore, if we calculate dimensionless measures of the time derivative of the Hubble rate we find contributions from perturbation theory like $(\partial^2\Phi)^2$, which are of order unity, and are actually divergent in the UV for a scale-invariant initial power spectrum; using any natural cutoff~-- from the end of inflation, say~-- gives colossal numbers. As we show here, these are not just a curiosity: we find the same thing if we calculate the monopole of the distance-redshift relation, a physical observable. We argue, as others have done~\cite{Kolb:2004am,Notari:2005xk,Kolb:2005da,Rasanen:2006kp,Rasanen:2010wz}, that it appears that higher-order perturbations will make things worse, not better. 

This is not to say that backreaction is necessarily large or significant, let alone responsible for an apparent acceleration effect.
Rather we argue that until we have a better grasp of the consequences of these divergences, and higher-order perturbation theory in general, it is impossible to conclude that backreaction is small. Clearly, linear perturbation theory is correct in many respects.  This success comes from analysing anisotropic and inhomogeneous perturbations, and observations of the CMB temperature anisotropy and the matter power spectrum, which all confirm it is basically correct. But backreaction concerns the homogeneous contribution of these perturbations at higher-order when they are calculated in terms of initial perturbations. Are they small?

\section{Measures of Backreaction}

In cosmology it is implicitly assumed that by statistical isotropy and homogeneity we simply construct a smooth FLRW `spacetime' from an lumpy inhomogeneous spacetime by averaging over structure. This might, in principle, have a metric 
\begin{equation}
\label{EffMetric}
\d s_{\mathrm{eff}}^{2}=-\d\tau^{2}+a_{\CD}^{2}\gamma_{ij}\d y^i\d y^j\mbox{ ,}
\end{equation}
where $\tau$ is the average cosmic time and $a_{\CD}(\tau)$ an averaged scale factor, the subscript $\CD$ indicating that it has been obtained at a certain spatial scale~$\CD$, which is large enough so that a homogeneity scale has been reached; in this case $\gamma_{ij}$ will be a metric of constant curvature.  Unfortunately, we don't know how to construct $\d s_{\mathrm{eff}}^{2}$ in an invariant way~-- there is no unambiguous averaging scheme. We don't know what field equations this metric would obey, nor do we know how to calculate observational relations; in fact we do not really know that any of the standard tools of general relativity survive.  There is no reason to believe any of these would be as in GR, as it has only been tested in the solar system where the averaging problem does not apply.

While different approaches to averaging exist which provide corrections to the field equations, those corrections require an underlying model with structure. In effect, it is as difficult to calculate a model with structure as the smooth average model which best corresponds to it. 
The scale factor itself comes from the solution to the field equations, but its first time derivative gives the Hubble expansion rate which is given algebraically from the Friedmann equation. Both of these change with averaging~\cite{Ell84,Buchert:1999er}.
We can estimate what the averaged scale factor in Eq.~(\ref{EffMetric}) might be by different properties of it directly within the standard model, using perturbation theory. Before we get to that, 
let us first consider measures of the expansion rate locally in a spacetime; then consider averaging spatially; and finally let us consider how averages of observables look in a general spacetime~-- which is what we actually use to infer properties of $a_\CD$.

\subsection{Exact spacetime dynamics}

Given a 4-velocity $u^a$, the local volume expansion is defined covariantly as 
\be
\Theta=3H=\nabla_a u^a=3u^a\nabla_a\ln \ell=3\frac{\dot\ell}{\ell}\,,
\ee
where $u^a$ is the local fluid 4-velocity with length scale $\ell$. Locally, the expansion rate obeys the usual Friedmann and Raychaudhuri equations:
\begin{eqnarray}
\Theta^2=3\left[\rho+ \Lambda + \sigma^2 -\sfrac{1}{2}{}^3\mathcal{R}\right]\\ 
\dot{\Theta} 
= - \,\sfrac{1}{3}\,\Theta^{2}- \sfrac{1}{2}\,(\rho+3p)+ \Lambda 
- 2\,\sigma^{2} + 2\,\omega^{2}
 + A_{a}A^{a} + \div A
\ ,
\end{eqnarray}
where the definitions of the terms are given in the Appendix.
For an exact solution to the field equations, the Friedmann equation is a first integral of the Raychaudhuri equation, along with other equations such as the evolution equation for the shear~-- see the Appendix. From this we can define a local deceleration parameter
\begin{eqnarray}
q_\Theta&=&-1-3\frac{\dot{\Theta}}{\Theta^2}, \\ \nonumber
&=&\frac{3}{\Theta^2}\left[\sfrac{1}{2}(\rho+3p)-\Lambda+2\sigma^2-2\omega^2-\div A-A_aA^a\right],\\
&=&\frac{\sfrac{1}{2}(\rho+3p)-\Lambda+2\sigma^2-2\omega^2-\div A-A_aA^a}{\rho+ \Lambda + \sigma^2 -\sfrac{1}{2}{}^3\mathcal{R}}
\end{eqnarray}
which is commonly used as a dimensionless measure of the rate of change of the expansion rate. Over a small interval of cosmic time both $\Theta$ and $q_\Theta$ give important information about the dynamics of $u^a$.

\subsection{Spatial averaging}

Let us assume for the moment an irrotational dust spacetime, and observers and coordinates at rest with respect to the
dust. The average of a scalar quantity $S$ may be (non-covariantly) defined as simply its integral
over a region of a spatial hypersurface $\mathcal{D}$ of constant
proper time divided by the Riemannian volume~(we follow \cite{Buchert:1999er}):
\begin{equation}
\label{eq:Average}
 \average{S(t,\bm{x})}=\frac{1}{V_\mathcal{D}}{\int_\mathcal{D}
 \sqrt{\det h}\, \d^{3}x \,\, S(t,\bm{x})}
\end{equation}
Taking the time derivative of Eq.~(\ref{eq:Average}) yields the commutation relation
\begin{equation}
\label{eq:CommRel}
\partial_t\average{S}-\average{\partial_tS}=\average{\Theta
S}-\average{\Theta}\average{S}\mbox{ ,}
\end{equation}
where $\Theta$ is the local expansion of the dust, and we assume the
domain is comoving with the dust. The dimensionless volume scale
factor is defined as $a_\mathcal{D}\propto {V_\mathcal{D}}^{1/3}$,
which ensures $\average{\Theta }=\partial_t\ln a_D$. Then, the equations of motion for this scale factor become:
\begin{eqnarray}\label{Av:Ray}
3\left(\frac{\dot{a}_{\mathcal{D}}}{a_{\mathcal{D}}}  \right)^2=\average{\rho}+\Lambda- \frac{1}{2}\left[\mathcal{Q}_\CD+\average{\mathcal{R}} \right]\\  \nonumber
3\frac{\ddot{a}_{\mathcal{D}}}{a_{\mathcal{D}}}  +\sfrac{1}{2}
\langle\rho\rangle_{\mathcal{D}}=\Lambda+ \mathcal{Q}_{\mathcal{D}},
\end{eqnarray}
where   $\mathcal{Q}_{\mathcal{D}}=
\frac{2}{3}\left[\langle\Theta^2 \rangle_{\mathcal{D}} - \langle
\Theta \rangle_{\mathcal{D}}^2\right]- 2\langle
\sigma^{2}\rangle_{\mathcal{D}} $ is the kinematic backreaction term and $\sigma^2 =\frac{1}{2}\sigma_{ab} \sigma^{ab}$ is the
magnitude of the shear tensor. The \emph{non-local} variance of the
\emph{local} expansion rate can act in the same way as the
cosmological constant, causing the average expansion rate to speed
up, even if the local expansion rate is slowing down. For consistency, and assuming $\average{\rho}\propto a_\CD^{-3}$, these equations must lead to the integrability condition:
\begin{equation}
\frac{1}{a_{\mathcal{D}}^6} \partial_t (\mathcal{Q}_{\mathcal{D}}a_{\mathcal{D}}^6) +\frac{1}{a_{\mathcal{D}}^2}\partial_t(\average{\mathcal{R}}a_{\mathcal{D}}^2)=0\,.
\end{equation}
A decceleration parameter $q_{\mathcal{D}}$ which describes the deceleration of the averaged hypersurface may be defined as:
\bea\label{qDdef}
q_{\mathcal{D}}&=& -\frac{1}{H_{\mathcal{D}}^2}\frac{\ddot a_{\mathcal{D}}}{a_{\mathcal{D}}}
\\
&=&\frac{\sfrac{1}{2}
\langle\rho\rangle_{\mathcal{D}}-\Lambda+2\langle
\sigma^{2}\rangle_{\mathcal{D}}- \frac{2}{3}\left[\langle\Theta^2 \rangle_{\mathcal{D}} - \langle
\Theta \rangle_{\mathcal{D}}^2\right] }
{\average{\rho}+\Lambda+ \langle
\sigma^{2}\rangle_{\mathcal{D}}-\frac{1}{2}\average{\mathcal{R}}
-\frac{1}{3}\left[\langle\Theta^2 \rangle_{\mathcal{D}} - \langle
\Theta \rangle_{\mathcal{D}}^2\right] }
\eea
  where $\frac{\ddot a_{\mathcal{D}}}{a_{\mathcal{D}}}$ is given by the averaged Raychaudhuri equation. However, it should be noted that this is very different from the \emph{average of the local deceleration parameter}, which is
\be\label{averagedecel_fluid}
\average{q_\Theta}=-1-3\bigg\langle\frac{\dot\Theta }{\Theta^2}\bigg\rangle_\CD.
\ee
As we can see, it is not easy nor unique to estimate $\ddot a_\CD$ given $a_\CD$, and we shall see that they are quite different in perturbation theory. 

These quantities come with a significant drawback: even for moderately sized domains, these quantities are unobservable because they are averaged on spatial hypersurfaces, whereas we only have observational access to our past lightcone (although see~\cite{Rasanen:2009uw}).

\subsection{Averaging physical observables}

Within the standard cosmology, cosmological parameters like the Hubble rate, and  deceleration parameter are well defined based on the background metric. These cosmological parameters can be evaluated today, at $t_0$, by taking a Taylor series expansion of the scale factor $a(t)$ 
\begin{eqnarray}
a(t)&=& a_0 \,
\left[1 + H_0 \,(t-t_0) + {1\over2} \,q_0 \,H_0^2 \,(t-t_0)^2 
+O([t-t_0]^3) \right].
\end{eqnarray}
Similarly, 
we can calculate  the important observational quantities of interest, such as the relation between the angular-diameter distance and redshift $d_A(z)$:
\begin{eqnarray}\label{FLRWlumdist}
d_A(z) = \frac{z}{H_0}\left(1-\frac{1}{2}\left(3+q_0\right)z+O(z)^2\right),\\
z = H_0 d_A+\frac{1}{2}\left(3+q_0\right)\left(d_AH_0\right)^2+O\left(d_AH_0\right)^3
\end{eqnarray}
Of course, these series expansions are only valid in an FLRW  background spacetime, but they can be generalised to an arbitrary spacetime using  the Kristain and Sachs approach~\cite{1966ApJ...143..379K,1969CMaPh..12..108E,1970CMaPh..19...31M}. This gives a way to define a generalised Hubble rate and deceleration parameter in an arbitrary spacetime, which are observationally rather than dynamically motivated; that these have important differences was first shown in~\cite{Clarksonthesis}.

The corresponding equations to Eqs.~(\ref{FLRWlumdist}) are:
\begin{eqnarray}\label{Gredshift}
z&=&[K^aK^b\nabla_bu_a]_0 d_A + \frac{1}{2}\left[K^aK^bK^c\nabla_a\nabla_bu_c\right]_0 d^2_A +\cdots\,,\\
\label{GAngdist}
d_A&=&\frac{z}{K^aK^b\nabla_au_b|_0}\left\{1-\left[\frac{1}{2}\frac{K^aK^bK^c\nabla_a\nabla_bu_c}{\left(K^dK^e\nabla_du_e\right)^2}\right]_0z+\cdots\right\}\,,
\label{Glumdist}
\end{eqnarray}
where the past pointing null vector is given by
\be
\label{K^a parallel and perpendicular}
K^a=\frac{k^a}{(u_bk^b)_0}=-u^a+e^a,
\ee
which points in the direction $e^a$ at the observer, located at `0'. Using the  notation whereby $K_{A_n}=K_{a_1}K_{a_2}\cdots K_{a_n}$ and $\nabla_{A_n}=\nabla_{a_1}\nabla_{a_2}\cdots \nabla_{a_n}$,
 the $n$'th term $K^{A_{n+1}}\nabla_{A_n}u_{a_n}$ for a given observer is an observable in principle, and will have a spherical harmonic expansion up to $\ell=n+1$. The monopole will be given by an all sky integral, $\int d\Omega K^{A_{n+1}}\nabla_{A_n}u_{a_{n+1}}$, and is usually what we would compare to observations in an FLRW context. 

In an arbitrary spacetime, the first terms in these expansions are given by
\bea
\fl
K^aK^b\nabla_au_b  = \sfrac{1}{3}\Theta -A_a e^a+ \sigma_{ab}e^ae^b\,
\\ \fl\nonumber
K^aK^bK^c\nabla_a\nabla_bu_c = \sfrac{1}{6}(\rho+3p) + \sfrac{1}{3}\Theta^2 -\sfrac{1}{3}\Lambda -\sfrac{2}{3}\omega_a\omega^a
+ \sigma_{ab}\sigma^{ab}+ A_aA^a-\sfrac{2}{3}\div A
\\ \nonumber   
- e^a\left[\sfrac{1}{3}\D_a\Theta+\sfrac{2}{5}\div{\sigma}_a +\dot A_a - \sfrac{4}{3} \Theta A_a - \sfrac{17}{5}A^b\sigma_{ab}\right]\\ \nonumber  
+e^{\<a}e^{b\>}\left[E_{ab}-\sfrac{1}{2}\pi_{ab}
+ \sfrac{5}{3}\Theta\sigma_{ab}+ \omega_{a}\omega_b + 2 {\sigma^c}_{a}\sigma_{bc}+ \epsilon_{acd}\omega^c{\sigma_{b}}^d-2\D_{a}A_{b}\right]
\\   
+ e^{\<a}e^be^{c\>}\left[A_{a}\sigma_{bc}- \D_{a}\sigma_{bc}\right]
\end{eqnarray}
The PSTF tensors $e^{\<A_\ell\>}$, for $\ell=0,1,2,3$, in this expression are a covariant representation of the spherical harmonics when evaluated at a given point in spacetime (see the appendix). \obinna{check signs}

In the same way that we may evaluate various aspects of the  spacetime by considering averages of the derivatives of $a_\CD$, we can investigate how observations might be affected by the evolution of structure. For example, the observable $K^aK^b\nabla_au_b$ gives us a generalised Hubble rate which now has a dipole and quadrupole in addition to the familiar monopole term~\cite{Clarksonthesis}. How shall we best compare higher order terms with the familiar FLRW expansion? Various definitions have been considered~\cite{Clarksonthesis,Barausse:2005nf}.

Note that the $d_A(z)$ relation has powers of $K^aK^b\nabla_au_b$ on the denominator of each coefficient. Since this means dividing by a spherical harmonic expansion, it will be much simpler to work with the $z(d_A)$ relation instead, and compare that with the FLRW relation. The definition we give below for the deceleration parameter is not unique, and depends on this choice. 

Let
\be
H^{\mathrm{obs}}_0=\sum_{\ell=0}^2\mathcal{H}_{A_\ell}e^{A_\ell}=[K^aK^b\nabla_au_b]_0
\ee
where the $\mathcal{H}_{A_\ell}$ are PSTF moments of the generalised Hubble rate $H^{\mathrm{obs}}_0$. Clearly the moments are simply, 
\be
\mathcal{H}=\frac{1}{3}\Theta,~~~\mathcal{H}_a=A_a~~\text{and}~~\mathcal{H}_{ab}=\sigma_{ab}\,, 
\ee
evaluate at the observer.
We now would like to relate the FLRW quantities $(3+q_0)H_0^2$ to $K^aK^bK^c\nabla_a\nabla_bu_c$, to define a generalised deceleration parameter. The simplest way is to define $q^{\mathrm{obs}}_0$ via:
\be
(3+q^{\mathrm{obs}}_0)\mathcal{H}^2=[K^aK^bK^c\nabla_a\nabla_bu_c]_0\,.
\ee
Then, writing $q^{\mathrm{obs}}_0=\sum_{\ell=0}^3 \mathcal{Q}_{A_\ell}e^{A_\ell}$, where the PSTF tensors $\mathcal{Q}_{A_\ell}$ are the multipole moments of the observational deceleration parameter, we have the monopole: \obinna{check the coeefs}
\be
\mathcal{Q}=\left.\frac{3}{\Theta^2}\left[\sfrac{1}{2}(\rho+3p)-\Lambda+6\sigma^2-2\omega^2-2\,\div A+3A^2\right]\right|_0\,.
\ee
Note that this is not the same as the local deceleration parameter associated with $\Theta$.

We have, then, three different ways to analyse the second-derivative of the expansion rate, all of which have been written as variants of the familiar deceleration parameter. We have the local exact deceleration parameter of which we can calculate the average over a domain, the deceleration parameter associated with the deceleration of the averaged hypersurfaces, and the `observed' deceleration parameter, which we have defined using the all-sky average (monopole) of the redshift-distance relation. In a general spacetime these are quite different things.

\section{Backreaction in the standard model}
\label{FLRW}

The standard model of cosmology ignores all the complexity of smoothing the spacetime by assuming that on `large' scales (say larger than a few hundred Mpc) we can model the universe as homogeneous and isotropic. Linear and higher-order fluctuations describing structure then propagate as smooth fields on this background. In theory, when we sum over all these smooth fields we should end up with a metric which describes the universe we see, with structure properly described on all relevant scales (statistically speaking).

We shall consider how backreaction comes about in the simplest cosmology which agrees with
observations: a flat LCDM model with Gaussian scalar
perturbations. In particular, we shall examine how backreaction affects the first and second derivatives of the scale factor, as given the Hubble rate and various deceleration parameters.
Averaging FLRW perturbations has been discussed often in the literature: some authors investigate specifically the modification to the Hubble expansion rate or other variables~\cite{1995ApJ...453..574Z,Russ:1996km,Boersma:1997yt,Rasanen:2003fy,Kolb:2004am,Barausse:2005nf,Li:2007ci,Li:2007ny,Li:2008yj,Kolb:2009rp,Clarkson:2009hr,Rasanen:2010wz,Umeh:2010pr}; others reformulate the average of the backreaction into an effective fluid~\cite{Noonan1984,Noonan1985,Paranjape:2006ww,Paranjape:2007wr,Paranjape:2007uj,Behrend:2007mf,Brown:2008ra, Baumann:2010tm,Chung:2010xx}, while one of  the first attempts considered the important problem of how to calculate the averaged metric~\cite{Stoeger:1999ig}. Many of these works consider only the case of backreaction in an Einstein-de Sitter model, in the hope of finding it responsible for dark energy. As it's more plausible that backreaction may lead to changes to the background at the level relevant for precision cosmology, we need to also investigate its effects in a LCDM model.

\subsection{Perturbation theory}

In the Poisson gauge to second-order in scalar perturbations the metric reads~
\begin{equation}
ds^2 = - \left[ 1 + 2 \Phi+\Phi^{(2)}\right] d t^2 - a
V_{i}dx^id t + a^2  \left [(1 - 2 \Phi-\Psi^{(2)})\gamma_{ij}+h_{ij} \right ]
dx^idx^j\,, \label{perturbed metric}
\end{equation}
The background evolution of the scale factor $a(t)$ at late times is determined by the Friedmann equation
\be
H(a)^2=\left(\frac{\dot a}{a}\right)^2=H_0^2\left[\Omega_ma^{-3}+1-\Omega_m\right]
\ee
where the Hubble constant $H_0$ is the present day expansion rate, and $\Omega_m$ the normalised matter content today.
The first-order scalar perturbations are given by
$\Phi, \Psi$ (and are all that is required for observations at the
moment), and the second-order by $\Phi^{(2)}, \Psi^{(2)}$ which
are needed for a consistent analysis of backreaction. We also include the second-order vector modes $V_i$ and tensor modes $h_{ij}$. In this
gauge we have the metric in its Newtonian-like form, which we may think
of as the local rest-frame of the gravitational field because it is the frame in which the
magnetic part of the Weyl tensor vanishes when vectors and tensors are ignored~\cite{Clarkson:2009hr}. 

For a single fluid with zero pressure and no anisotropic stress
$\Psi=\Phi$, and $\Phi$ obeys the `master' equation
\begin{equation}
\ddot\Phi+4H\dot\Phi+\Lambda\Phi=0\,.
\label{eq:BE}
\end{equation}
For a LCDM universe the solution in time to this is equation has $\Phi$ constant until $\Lambda$ becomes important, and then starts to decay as $\Lambda$ suppresses the growth of structure on all scales by about a factor of 2. We write it as $\Phi(t,\bm x)=g(t)\Phi_0(\bm x)$ where $g(t)$ is the growing solution to Eq.~(\ref{eq:BE}) normalised to $g=1$ today (we can use $g_\infty=g(t=0)\approx \frac{1}{5}(3+2\Omega_m^{-0.45})$ as a very good approximation to its early time value). There is no scale dependence in the equation, which all comes from the initial conditions~-- usually a nearly scale-invariant Gaussian spectrum from frozen quantum fluctuations during inflation~-- and subsequent evolution during the radiation era. Evolution during the radiation era suppresses wavelengths which enter the Hubble radius compared to those which remain larger than it until the matter era begins. 
Consequently, in Fourier space, assuming scale invariant initial conditions from inflation, the power spectrum of $\Phi$, ${\cal P}_\Phi$, is independent of scale for modes larger than the equality scale, $k_{eq}=\sqrt{2\Omega_m z_{eq}}H_0\approx0.07\Omega_mh^2$\,Mpc$^{-1}$. A dimensionless transfer function describes the loss of power in the case of zero baryons (adapted from~\cite{Eisenstein:1997ik}):
\be
T(k)=\frac{\ln  \left(  2e+ 0.134\,\kappa \right)}{   \ln  \left(  2e+
 0.134\,\kappa \right) + \left[  0.079+ \frac{4.06}{ 1+
 4.66\,\kappa } \right] {\kappa}^{2}}\,
\ee
where $\kappa=k/k_{eq}$. This is unity for $\kappa\ll1$ and $\sim(\ln\kappa)/\kappa^2$ for $\kappa\gg1$.
The change in behaviour at the equality scale is important for backreaction because it is the modes larger than the equality scale which are primarily responsible for any backreaction at all. In essence, the equality scale determines the size of the backreaction effect.

All first-order quantities can be derived from $\Phi$; for
example, 
\be
v^{(1)}_i=-\frac{2}{3aH^2\Omega_m}\partial_i\left(\dot\Phi+H\Phi\right), 
\ee
is the first-order velocity perturbation, which governs the peculiar
velocity between the matter flow and the rest-frame of the
gravitational field. Meanwhile, the gauge-invariant density perturbation is
\be
\delta=\frac{\delta\rho}{\rho}=\frac{2}{3H^2\Omega_m}\left[a^{-2}\partial^2\Phi-3H\left(\dot\Phi+H\Phi\right)\right]\,.
\ee

The second-order solutions for $\Psi^{(2)}$ and $\Phi^{(2)}$ are
given by~\cite{Bartolo:2005kv}. These are complicated expressions involving
time integrals over products of $\Phi$ and its derivatives. For
backreaction, however, the important thing is that if we take into account only terms which are important for backreaction we have simply
\be
\Psi^{(2)}=\Phi^{(2)}= B_3(t) \partial^{-2} \partial_i\partial^j(\partial^i \Phi_0 \partial_j \Phi_0 )
+B_4(t) \partial^i \Phi_0 \partial _i\Phi_0\,,
\ee
where $B_3(t)$ and $B_4(t)$ are time dependent functions given by integrals over $g(t)$, with dimension $H_0^{-2}$. Their values today are given roughly by $B_3(t_0)\sim0.85/(\Omega_m^{1.15} H_0^2)$ and $B_4(t_0)\sim-0.25/(\Omega_m^{1.15} H_0^2)$. $\Psi^{(2)}$ and $\Phi^{(2)}$ also contain more complicated looking inverse Laplacian terms, but these drop out after ensemble averaging~\cite{Clarkson:2009hr}, so we ignore them here. We shall only explicitly calculate scalar quantities at second-order here, so we don't require the exact form of the vectors and tensors; for reference, generically we have~\cite{Mollerach:2003nq,Ananda:2006af,Baumann:2007zm,Lu:2008ju,Lu:2008ju}:
\be
V_i\sim\Phi\partial_i\Phi,~~~h_{ij}\sim \Phi\partial_{\<i}\partial_{j\>}\Phi\,,
\ee
which we shall return to later.

\subsection{Backreaction of perturbations}

What is the backreaction of perturbations onto the the expansion rate and its associated deceleration parameter? Similarly, what are the corrections to the monopole of the distance-redshift relation from second-order perturbations?

Expressions are quite cumbersome at second-order, so let us define the Hubble normalised dimensionless derivative 
\be
\tilde{\partial}=(aH)^{-1}\partial\,.
\ee
At second-order, the expansion rate of the dust is 
\begin{eqnarray}
\frac{\Theta}{3H}&=&1 - (1+\hat{g})\Phi 
-\sfrac{2}{9}\left(1+\hat{g}\right)\tilde\partial^2\Phi\nonumber\\ && 
+ \sfrac{1}{2}\left(3- 2\hat{g}\right)\Phi^2
-\sfrac{1}{2}\Phi^{(2)}-\sfrac{1}{2}H^{-1}\dot{\Psi}^{(2)}
+ \sfrac{1}{6}a\tilde\partial_k v_{(2)}^k \nonumber\\ && 
+\frac{2}{27\Omega_m^2}\left[\left(1+2\hat{g} +\hat{g}^2\right)+9\Omega_m \left(1+\hat{g}\right)\right]\tilde\partial_k\Phi\tilde\partial^k\Phi\\
&\Rightarrow& \mathcal{H}~\text{when evaluated today at the observer}\,.\nonumber
\end{eqnarray}
This quantity is also the sky-averaged Hubble rate, measured by dust observers from observations on the past nullcone. We have used the fact that we are only interested in the growing mode solutions for $\Phi$, and have written $\dot\Phi=\dot g\Phi_0=\hat g H\Phi$ where $\hat g=\dot g/gH$ (a good approximation to its value today is $\hat g\approx-1.31(1-\Omega_m^{0.4})$).
In principle, then, the perturbative terms give the backreaction to the local expansion rate. Here, the divergence of the second-order velocity terms is~\obinna{check}
\begin{eqnarray}
\fl a\Omega_m^2 \tilde\partial_k v_{(2)}^k= -\frac{4}{3}H\left((2-\Omega_m)+ 4 \hat{g} + 2  \hat{g}^2\right)\left(\Phi\tilde\partial^2\Phi+ \tilde\partial_k\Phi\tilde\partial^k\Phi\right)\\  
- \frac{2}{3}\Omega_m  \left(H\tilde\partial^2\Phi^{(2)}+ \tilde\partial^2\dot{\Psi}^{(2)}\right)
 + \frac{8}{9}\left(1+ \hat{g}\right)\left(\tilde\partial^2\Phi \tilde\partial^2\Phi+ \tilde\partial_k \tilde\partial^2\Phi \tilde\partial^k\Phi\right)\,.\nonumber
\end{eqnarray}
Note the terms with four derivatives in them~-- individually these are large, but they cancel out after ensemble averaging as we discuss below (or, in the language of~\cite{Rasanen:2003fy}, they only constitute a boundary term so should vanish on large scales).

We may also calculate two measures of the deceleration within perturbation theory directly as they don't rely on averaging: $q_\Theta$ and $\mathcal{Q}$. These are rather messy, but have the same form up to coefficients. 

For example, the monopole of the deceleration parameter defined through the distance-redshift relation becomes
\begin{eqnarray}\fl
\mathcal{Q} =  -1+ \frac{3 }{2}\Omega_m - 3(1+\hat{g})(1-\Omega_m)\Phi+ \left[(1+\sfrac{2}{3}\hat{g})-\frac{4}{9\Omega_m}(1+\hat{g})\right]\tilde{\partial}^2\Phi \nonumber 
\\ \nonumber  
-3\hat{g}\left(4+\sfrac{3}{2}\hat{g}\right)(1-\Omega_m)\,\Phi^2
-\frac{3}{2H}(1-\Omega_m)(H\Phi^{(2)}+\dot{\Psi}^{(2)})
\\ \nonumber  
- \sfrac{1}{2}{a\Omega_m\tilde{\partial}_kv_{(2)}^k}- \frac{a}{3H}{\tilde{\partial}_k\dot{v}_{(2)}^k} 
-\sfrac{1}{3}\tilde{\partial}^2\Phi^{(2)}+\sfrac{1}{6}\tilde{\partial}^2\Psi^{(2)}
\\ \nonumber  
+ \frac{1}{9\Omega_m}\left[3\Omega_m(10+14\hat{g}+6\hat{g}^2)-8(2+5\hat{g}+3\hat{g}^2)\right]\Phi\tilde{\partial}^2\Phi 
\\ \nonumber  
-\frac{1}{\Omega^2_m}\left[\sfrac{1}{6}\Omega_m^2(13+12\hat{g})-\sfrac{4}{9}\Omega_m(2-\hat{g}-12\hat{g}^2) -\sfrac{4}{27}(1+2\hat{g}+\hat{g}^2)\right]\tilde{\partial}_k\Phi\tilde{\partial}^k\Phi
\\ \nonumber  
+\frac{4}{27\Omega_m^2}(1+2\hat{g}+\hat{g}^2)\tilde{\partial}_i\tilde{\partial}_j\Phi\tilde{\partial}^i\tilde{\partial}^j\Phi
-\frac{8}{27\Omega_m^2}(1+2\hat{g}+\hat{g}^2)\tilde{\partial}^k\Phi\tilde{\partial}^2\tilde{\partial}_k\Phi
\\ \nonumber  
+\frac{2}{27 \Omega_m}\left[\Omega_m(5+8\hat{g}+3\hat{g}^2)-4(1+2\hat{g}+\hat{g}^2)\right]\tilde{\partial}^2\Phi\tilde{\partial}^2\Phi
\end{eqnarray}
where it is understood that the rhs is evaluated at the observer, and is not a spacetime function in the usual sense. The other deceleration parameters are given in the appendix.

\subsection{Spatial Averaging}

Consider now the average of a variable over a spatial domain $\cal D$. The Riemannian average of a quantity $\Upsilon$ may be defined as,
\be
\langle \Upsilon\rangle_{\CD}=\frac{1}{V_{\CD}}\int_\CD \sqrt{\det h}\, \d^3 x \Upsilon
\ee
Here, the domain is on a hypersurface with 3-metric $h_{ij}$. If we choose that hypersurface to coincide with the spatial surfaces of the metric in the longitudinal gauge~-- the gravitational rest-frame~-- then this
 can be easily expanded in terms of the Euclidean average  defined on the background space slices,
$\langle \Upsilon \rangle ={\displaystyle\int_\CD\d^3x\, \Upsilon }\bigg/{\displaystyle\int_\CD \d^{3}x}$, as:
\begin{equation}
\label{ExpandAverage} \saverage{\Upsilon}_\text{grav}=
\Upsilon^{(0)}+\langle\Upsilon^{(1)}\rangle+\langle\Upsilon^{(2)}\rangle+
3\left[\langle\Upsilon^{(1)}\rangle\langle\Psi\rangle-\langle\Upsilon^{(1)}\Psi\rangle\right]\mbox{,}
\end{equation}
where $\Upsilon^{(0)}$, $\Upsilon^{(1)}$ and $\Upsilon^{(2)}$
denote respectively the background, first order and second order
parts of the scalar function
$\Upsilon=\Upsilon^{(0)}+\Upsilon^{(1)}+\Upsilon^{(2)}$. Note the important term in square brackets, which encapsulates the relativistic part of the averaging procedure. 
To link with Buchert's formulation above  we would like to average in the hypersurfaces orthogonal to $u^a$~-- the local rest frame of the dust~-- in which case we have instead
\begin{eqnarray}
\fl\langle \Upsilon \rangle_{\CD}= \Upsilon^{(0)} +\saverage{\Upsilon^{(1)}} -g_I\dot{\Upsilon}^{(0)}\saverage{\Phi}
+\saverage{\Upsilon^{(2)}}\\
+(1-3Hg_I)\left[\saverage{\Upsilon^{(1)}\Phi}-\saverage{\Upsilon^{(1)}}\saverage{\Phi}\right]-g_I\saverage{\Phi \dot{\Upsilon}^{(1)}}\nonumber\\ 
+g_I^2\left[3H{\dot{\Upsilon}^{(0)2}}+\frac{1}{2}\ddot{\Upsilon}^{(0)}\right]\saverage{\Phi^2}
+ 3g_I(1-g_IH)\dot{\Upsilon}^{(0)}\saverage{\Phi}^2 \nonumber\\
-\frac{1}{2}\dot{\Upsilon}^{(0)}\int^t\d t' \left[
\saverage{\Phi^{(2)}}-\saverage{\Phi^2}-2\saverage{v^{(1)}_i v_{(1)}^i}-2g_Ia^{-2}\saverage{v_{(1)}^i\partial_i \Phi}\right]\nonumber
\end{eqnarray}
 where $ g_I= \frac{1}{g(t)}\int^t g(t')\d t'$. So, for example, to find $\average{q}$ use $\Upsilon^{(0)}=-1+ \frac{3}{2}\Omega_m$, $\Upsilon^{(1)}=- 3 (1+\hat{g})(1-\Omega_m)\,\Phi 
+\frac{1}{9\Omega_m }\left[(9+6\hat{g})
-4\Omega_m(1+\hat{g})\right]\,\tilde\partial^2\Phi$ and $\Upsilon^{(2)}= \cdots$~all the other stuff~$\cdots$,  in this expression (cf the Appendix).

With this we now have 
\bea\fl
\<\Theta\>_\mathcal{D}=\text{the average of the expansion rate, averaged in the fluid rest frame}\nonumber\\\fl
\<q_\Theta\>_\mathcal{D}=\text{the average of the deceleration parameter, averaged in the fluid rest frame}\nonumber\\\fl
q_\CD=\text{the deceleration of the scale factor associated with~}\<\Theta\>_\mathcal{D}\nonumber\\\fl
\mathcal{Q}=\text{the monopole of the locally observed deceleration parameter}
\eea
We now turn to estimating the amplitude of these quantities given standard initial conditions from inflation.

\subsection{Ensemble Averaging}

We define our Fourier transform as (suppressing any temporal quantities)
\be
\Phi(\bm x)=\frac{1}{(2\pi)^{3/2}}\int \d^3k\, \Phi(\bm k)\, e^{i\bm k\cdot\bm x}, 
\ee
where $\Phi^*(\bm k)=\Phi(-\bm k)$. Note that we assume $\Phi(\bm x)$ is defined over an infinite volume for this definition to be valid, as it requires the crucial identity $(2\pi)^3\delta(\bm x)=\int\d^3k e^{i\bm k\cdot\bm x}$. (This can instead be reformulated in a large box with periodic boundary conditions~\cite{Lyth:2007jh}. The periodicity of the boundary conditions is instead contained in $\delta$-functions at $\bm k=0$ here.)

Inflationary models typically provide us with initial conditions for $\Phi$ in terms of its correlations, which appear in the form of an ensemble average or expectation value: for example, the 2-point correlation function is just
\be
C(\bm k,\bm k')=\overline{\Phi(\bm k)\Phi(\bm k')}\,,
\ee
where an over-bar denotes an ensemble average.
If $\Phi$ is Gaussian, then the distribution is given entirely by $C(\bm k,\bm k')$, and higher correlations are given in terms of $C$ using Wick's theorem, and the ensemble average of odd numbers of $\Phi$ is zero. The probability distribution function is just $P[\Phi(\bm k)\Phi(\bm k')]\sim\exp\left[\frac{\Phi(\bm k)\Phi(\bm k')}{2C(\bm k,\bm k')}\right]$\,. Statistical homogeneity of $\Phi(\bm x)$ implies that different modes are uncorrelated: $C(\bm k,\bm k')\propto \delta(\bm k+\bm k')$; statistical isotropy implies that the proportionality function cannot depend on the direction of~$\bm k$~\cite{Abramo:2010gk}. Hence we have our power spectrum:
\be
\overline{\Phi(\bm k)\Phi(\bm k')}=\frac{2\pi^2}{k^3}\mathcal{P}_\Phi(k)\delta(\bm k+\bm k')\,. 
\ee
Assuming scale-invariant initial conditions from inflation, this is given by
\be
\mathcal{P}_\Phi(t,k)=\left( \frac{3 \Delta_\mathcal{R}}{5
g_{\infty}} \right )^2 g(t)^2 T(k)^2 \,, 
\ee
where $\Delta_{\mathcal{R}}^{2}$ is the primordial power of the
curvature perturbation~\cite{Komatsu:2008hk}, with
$\Delta_{\mathcal{R}}^{2} \approx 2.41 \times 10^{-9}$ at a scale
$k_{CMB}=0.002 \mathrm{Mpc}^{-1}$.

What is the relation between ensemble and spatial averaging? Usually we assume ergodicity which roughly means here that, because the space is assumed infinite, the ensemble can be found within the space itself~-- i.e., any realisation exists somewhere. Consequently we should be able to replace ensemble averages with spatial ones, and vice versa. This only works if the spatial averages are taken over an infinite domain \emph{on the background}, however, which is not necessarily what we are interested in when considering the averaging problem. We would like to know how properly averaged quantities change with scale. To see this, define the Euclidean average over a domain $\CD$ using a window function $W(\bm x-\bm x')$. Then
\be
\saverage{X(\bm x')}(\bm x)=\frac{1}{V}\int\d^3 x' W(\bm x-\bm x') X(\bm x')\,,
\ee
where $V(\bm x)=\int d^3 x' W(\bm x-\bm x')$ gives the volume associated with $W$. For two correlated Gaussian random fields $A$ and $B$ satisfying $\overline{A(\bm k)B(\bm k')}=A(k)B(k)\delta(\bm k+\bm k')$ we have $\overline{A(\bm x)B(\bm x)}=(2\pi)^{-3}\int\d^3 kA(k)B(k)$ directly. However, when averaged over a finite domain, we have instead:
\bea
\saverage{A(\bm x')B(\bm x')}(\bm x)&=&\frac{1}{V}\int\d^3 x' W(\bm x-\bm x') A(\bm x')B(\bm x')\nonumber\\
&=& \frac{1}{(2\pi)^3}\int\d^3 k_1\int \d^3 k_2 W(\bm k_1+\bm k_2\,;\,\bm x) A(\bm k_1)B(\bm k_2)\nonumber
\eea
where $W(\bm k\,;\,\bm x)=V(\bm x)^{-1}\int\d^3 x'W(\bm x-\bm x') e^{-i\bm k\cdot\bm x'}$ is the Fourier transform of $W/V$ anchored at $\bm x$. In the limit we average over all space,  $W/V\to 1$ then $W(\bm k)=(2\pi)^3\delta(\bm k)$ and is independent of the anchor point $\bm x$. 
Clearly we recover the ensemble averaged result in this case, provided the fields are isotropic. \obinna{there are still 2pi factors wrong here...}

\subsection{The size of the backreaction}

To evaluate our various averaged quantities we could use a realisation of $\Phi$ given an inflationary model.  Alternatively, we can assume a spectrum for $\Phi$ and evaluate the statistics of the quantity in question. This allows us to calculate the expectation value of averaged variables as well as their ensemble variance, in terms of integrals over the power spectrum of $\Phi$ multiplied by powers of $k$. The reason we must go to second-order now becomes clear when we calculate the expectation value of an average: for Gaussian perturbations from inflation, the ensemble average of $\Phi$ is zero, which implies~-- assuming ergodicity~-- that when averaged on the background over a very large (strictly, infinite) domain they are zero too. 
Thus, the second-order terms provide the principal backreaction effect; the first-order terms give the statistical variance.

Let us estimate the approximate behaviour of each type of terms which appears. 
The relations for determining the scaling behaviour for the backreaction terms are ($n+m$ is even) 
\bea\label{av-phi}
\overline{{\tilde\partial^m\Phi\,\tilde\partial^n\Phi}}
&=&\frac{(-1)^{(m+3n)/2}}{(aH)^{n+m}}\int_{0}^\infty \d k\, k^{m+n-1}\mathcal{P}_\Phi(k)\,.
\eea
The inverse Laplacian term in $\Phi^{(2)}$ satisfies $\overline{\saverage{\partial^{-2} \partial_i\partial^j(\partial^i \Phi_0 \partial_j \Phi_0 )}}=\sfrac{1}{3}\overline{\saverage{\partial^i \Phi_0 \partial_j \Phi_0}}$~\cite{Clarkson:2009hr}. 
We also have that, since $\Phi$ is statistically homogeneous and isotropic~\cite{Clarkson:2009hr} 
\be
\overline{\partial^2\Phi^{(2)}}=\overline{\partial^2\Psi^{(2)}}=0,~~~\overline{\partial_k v^k_{(2)}}=0\,,
\ee
which means that all the potentially large terms in the second-order Hubble rate don't contribute to the expectation value (see below).  For the non-connected terms we have
\be
\overline{\saverage{\tilde\partial^m\Phi}\saverage{\tilde\partial^n\Phi}}
=\frac{(-1)^{(m+3n)/2}}{(aH)^{n+m}}\int_{0}^\infty \d k\, k^{m+n-1}W(kR_D)^2\mathcal{P}_\Phi(k),
\ee
where $W$ is an appropriate window function specifying the domain. Typically this will become a delta-function as the domain tends to infinity (e.g., if the window function in real space is a Gaussian of width $R_\CD$, then in Fourier space it is a Gaussian of width $1/R_\CD$, centred at $k=0$). Note that the connected terms have no dependence on the domain size or shape at all, and that the domain dependence arrises from the non-connected terms~-- these in turn come from using the Riemannian volume element. The integral can be written as
\be\label{eq49}
\frac{1}{H^{n+m}}\int_{0}^\infty \d k\, k^{m+n-1}\mathcal{P}_\Phi(k)=
\left( \frac{3 \Delta_\mathcal{R}}{5
g_{\infty}} \right )^2\left(\frac{k_{eq}}{k_H}\right)^{m+n}\int_0^\infty\d\kappa\, \kappa^{m+n-1}T(\kappa)^2\,.
\ee
Here, $k_H=H_0^{-1}$ is the wavenumber of the mode entering the Hubble rate today, and $k_{eq}/k_H=\sqrt{2\Omega_m z_{eq}}\sim40$. Using
$z_{eq}\approx 2.4\times10^4\Omega_m h^2$ implies the important relation 
\be\label{magic}
\Delta_\mathcal{R}\left(\frac{k_{eq}}{k_H}\right)^2\approx 2.4\, \Omega_m^2 h^2\,.
\ee
For pure CDM and a scale-invariant initial spectrum, the integral behaves as, replacing $\int_{0}^\infty\mapsto\int_{\kappa_\text{IR}}^{\kappa_\text{UV}}$ where necessary:
\be
\int_0^\infty\d\kappa\, \kappa^{m+n-1}T(\kappa)^2\approx
\left\{
\begin{array}{ccc}
-\ln(\kappa_\text{IR})  & \text{for}  & m+n=0  \\
3.9  & \text{for}  & m+n=2  \\
F(\kappa_\text{UV})  & \text{for}  & m+n=4  
\end{array}
\right.
\ee
The function $F$ is roughly $F(x)\sim 0.44\, x^{2.14}$ for $1\lesssim x\lesssim 10$, $\sim 70 x^{-0.1}(\log_{10}x)^{4.75}$ for $x\gg1$, and approaches $\sim 53\ln^3 x$ as $x\to\infty$. For integrals with the window function inside, $W(\kappa k_{eq} R_\CD)$, we can roughly replace $\kappa_\text{UV}\mapsto  1/R_\CD k_{eq}$, though this depends on the details of the window function used. Combining the above equations allows us to calculate reasonably precisely the size of each term at second-order. 


The first type of term, $\Phi^2$, is nominally tiny,~$\mathcal{O}(10^{-10})$. It also tells us that the IR divergence in $\Phi^2$ must be cut-off by hand, corresponding to the first mode to leave the Hubble radius at the start of inflation. There has been speculation that this might lead to very important effects and even mimic dark energy~\cite{Kolb:2004am,Kolb:2005da,Barausse:2005nf}, though this has been criticised~\cite{Flanagan:2005dk,Hirata:2005ei,Geshnizjani:2005ce}. Given that we only measure the primordial power spectrum to be nearly scale-invariant over a comparatively narrow range of scales, the appearance of $\overline{\saverage{\Phi^2}}$ implies that it can't be too tilted to the red on super-Hubble scales, or else it would give a sizeable backreaction effect, though this might be unphysical. A red spectrum would convert the logarithmic divergence into a power-law one, and constraints on the largest mode would be much stronger (e.g., $k_{IR}/k_{H}\lesssim 10^{\sim80}$ for $n_s=0.95$).

The term primarily responsible for setting the fundamental amplitude of the backreaction in the Hubble rate is $\overline{\saverage{\Phi\tilde\partial^2\Phi}}\propto (k_{eq}/k_H)^2$. It is nominally quite small, 
\be
\frac{\overline{\saverage{\Phi\partial^2\Phi}}}{\Omega_mH_0^2}\sim  \Delta_{\cal R}^2 \frac{k_{eq}^2}{\Omega_m k_H^2}\sim  \Delta_{\cal R}^2 \frac{T_{eq}}{T_0}\sim 10^{-5}
\ee
for the concordance model. (The overall effect is somewhat larger than this due to the contribution of several such terms.) This gives sub-percent changes to the Hubble rate from backreaction, though non-connected terms make it significantly larger on small scales~\cite{Li:2007ci}. Yet, we can observe that the \emph{backreaction is small because the equality scale is large in our universe}, which is because the temperature of matter-radiation equality is very low. Modes which enter the Hubble radius during the radiation era are significantly damped compared to those which remain outside until after equality; so, the longer the radiation era, the less power there is on small scales to cause a significant backreaction effect.  
For a scale-invariant spectrum, then, it may be considered that the  long-lived radiation era is the reason that the dynamical backreaction is small. The temperature will have to drop by several orders-of-magnitude before backreaction in the Hubble rate becomes significant.

In the variance of the Hubble rate, the deceleration parameter $q_\Theta$, and the ensemble average of $\mathcal{Q}$ divergent terms appear: ${\tilde\partial^2\Phi}{\tilde\partial^2\Phi}$, which are of order the density fluctuation squared\footnote{While they appear in the Hubble rate through the second-order velocity perturbation, the ensemble average conspires to cancel them out for an initial spectrum which is homogeneous. Because the peculiar velocity appears as a pure divergence, the spatial average is a boundary term which is small on infinite domains due to assumed homogeneity~-- in our analysis this is contained in assuming ergodicity of the first-order $\Phi$.}. The origin of such terms arises from taking the \emph{proper} time derivative of the Hubble rate which contains spatial derivatives with respect to the coordinates we are using.  The integral $F$ overcomes the $\sim10^{-9}\times (2\Omega_m z_{eq})^2$ pre-factor around $k_{UV}\sim 10k_{eq}$, so these terms are big, and difficult to know what to do with: here, the UV cutoff is really a measure of our ignorance. Within linear perturbation theory it should be set by the end of inflation and the reheating temperature, as well as the small-scale physics of dark matter, both of which are sub-pc scales today. Replacing the UV cutoff with a smoothing function in $\Phi$ implies that we might do better in perturbation theory to smooth order-by-order, and calculate second-order terms from smoothed first-order ones, rather than calculating expectation values or spatial averages directly at a given order. 
Even for domains much larger than the non-linear scale, where linear perturbation theory breaks down (somewhere around a few Mpc), $F$ is quite sizeable, and so we have backreaction terms ${\cal O}(1)$ (note that $F(1)=\mathcal{O}(1)$, and the prefactor is also $\mathcal{O}(1)$). From this, we also recover that the variance in the Hubble rate is ${\cal O}(1)$ on scales of Mpcs. 

Observationally, can there be any signature of backreaction? When measuring the Hubble rate, perturbations are significant in the variance of the Hubble rate on sub-equality scales~\cite{Li:2007ci,Li:2008yj}. Perturbations affect the whole distance-redshift relation which has been calculated to first-order by~\cite{Bonvin:2005ps,Bonvin:2006en,Barausse:2005nf}. Corrections to the luminosity distance include corrections $\partial^2\Phi$, just like for the Hubble rate. This allows us to see that the variance includes divergent terms like $(\partial^2\Phi)^2$~-- the lensing term~-- which appear in the variance of the all-sky average of the luminosity distance~-- see \cite{Bonvin:2005ps}. We have discussed here how the problem at second-order is much more serious: the divergent terms appear in the expectation value of the distance-redshift relation itself, and not just in its variance.  

Consider the ensemble average of the monopole of $q^\text{obs}_0$. This should be the value a typical observer would expect to find~-- from nearby supernovae for example. Considering only the dominant terms with 4 derivatives in them, and using 
$\Delta_\mathcal{R}^2\left({k_{eq}}/{k_H}\right)^4\approx 5.7\, \Omega_m^4 h^4,
$ we have $\mathcal{O}(1)$ corrections to the observed deceleration parameter:
\bea\fl
\overline{\mathcal{Q}}
\approx-1+\frac{3}{2}\Omega_m +\Omega_m^4 h^4 \left[1.06+0.03(1-\Omega_m)^2-1.4(1-\Omega_m)^{11}\right] F(\kappa_\text{UV})\\
\fl\text{similarly\,,}\nonumber\\
\fl
\overline{q_\Theta}\approx-1+\frac{3}{2}\Omega_m +\Omega_m^4 h^4 \left[-0.24(1-\Omega_m)^{3.94}+0.66\Omega_m^{0.37}\right] F(\kappa_\text{UV})
\eea
where the coefficient in square brackets are reasonable empirical estimates for $\Omega_m\gtrsim0.1$ (accurate to a \% or so). In Buchert's definition of $q_\mathcal{D}$, the divergence is nicely controlled by the domain size because terms $(\partial^2\Phi)^2$ always appear as $\average{\partial^2\Phi}^2$. In that case the UV cutoff can be thought of as the smoothing scale.

It is an important open question to find out what happens to the whole distance-redshift relation to second-order, where an overall ensemble averaged offset to the luminosity distance will be present. As we have shown, this will include terms $(\partial^2\Phi)^2$, which is certainly important enough to disturb the cosmic concordance.

\subsection{Higher-order perturbation theory}

Would higher-order perturbation theory affect these conclusions? On the one hand it seems clear that higher-order perturbations should be suppressed. Provided $\Phi$ is Gaussian, only even orders will be important, once ensemble averages are taken. We might expect the largest terms at any order $n$ to behave like $\Phi^{(n)}\sim(\partial\Phi)^n$ (e.g., from relativistic corrections to the peculiar velocity), the ensemble average of which goes like $\Delta_{\cal R}^n (k_{eq}/k_H)^n$. Terms which appear in the Hubble rate at order $n$ of the form $\partial^2\Phi^{(n)}$ do not have enough derivatives to overcome the suppression from $(\partial\Phi)^{(n-2)}$ terms. By this argument, second-order should be as large as it gets, and backreaction from structure is irrelevant. 

On the other hand, others~\cite{Kolb:2004am,Notari:2005xk,Kolb:2005da,Rasanen:2006kp,Rasanen:2010wz} have argued that at higher order terms such as $(\partial\Phi)^2(\partial^2\Phi)^{n-2}$ are the norm. Simply squaring terms like $\Phi\partial^2\Phi$ gives problematic terms. In this case, from 4th order on perturbation theory needs a UV cutoff~-- at least as far as calculating averages is concerned. Even if not divergent, if $(\partial^2\Phi)^{n-2}\sim 1$ then higher-order terms are at least as large as at second-order and must be included to evaluate backreaction properly, and so correctly identify the background. But do these terms cancel out? 

Let us consider the expansion rate at fourth order, as an example. If we are interested in the expectation value, the leading terms, which contain 6 or more derivatives, are: 
\begin{eqnarray}\fl
\frac{H^{(4)}}{H}\sim - \frac{16}{729\Omega_m^4}\left(1+3\hat{g}+3\hat{g}^2+\hat{g}^3\right)\,  \tilde\partial_k\Phi\tilde\partial^k\Phi\tilde\partial_j\tilde\partial^2\Phi\tilde\partial^j\Phi\\ \nonumber
+ \frac{8}{243\Omega^4_m}\left[3(1-\hat{g}^2-2\hat{g}^3)+\Omega_m(27-67\hat{g}+30\hat{g}^2)\right]\, \tilde\partial_k\Phi \tilde\partial^k\Phi\tilde\partial^2\Phi\tilde\partial^2\Phi\\ \nonumber
-\frac{16}{243\Omega^4_m}\left[2(1+3\hat{g}+3\hat{g}^2+\hat{g}^3)+7\Omega_m(1+\hat{g})\right]\,\tilde\partial_k\Phi\tilde\partial_k\tilde\partial_j\Phi\tilde\partial^j\Phi\tilde\partial^2\Phi\nonumber\\
+\frac{1}{81\Omega_m^3}\left(\tilde\partial_k\tilde\partial^2\Psi^{(2)}-\tilde\partial_k\tilde\partial^2\Phi^{(2)}\right)\left(H\tilde\partial^k\Phi^{(2)}+\tilde\partial^k\dot\Psi^{(2)}\right)
\nonumber\\
\fl+\text{~terms of the form:~~}\Phi^{(4)},~\partial_k v^k_{(4)},~\partial_k\Phi\partial^k\Phi\partial^2 \Phi^{(2)},\nonumber\\
\partial_k\partial^2 \Phi^{(3)}\partial^k\Phi,~~ \partial^2\Phi\partial^i\Phi\partial_i\Phi^{(2)},~~\partial_k \Phi^{(2)}\partial^k \Phi^{(2)},\ldots\nonumber
\end{eqnarray}
Here, we have assumed the metric can be written in the longitudinal gauge up to $\mathcal{O}(4)$, and we have ignored all stand-alone cubic terms as they do not appear in the ensemble average. We have also ignored coupling with vector and tensor modes which will give potentially important contributions. We have also grouped $\Phi^{(2)}$ and $\Psi^{(2)}$ together, schematically unless explicitly written out. Note that there is the potential for 8 derivative terms from terms like  $(\partial_k\partial^2 \Phi^{(2)}-\partial_k\partial^2 \Psi^{(2)})\partial^k \Phi^{(2)}$ which appear. However, on closer inspection they cancel on small scales where $\Phi^{(2)}\simeq\Psi^{(2)}$, so should not dominate the expression. We have included the coefficients where it is obvious how to calculate the ensemble average using Wick's theorem; where terms include $\Phi^{(2)}$ and $\Phi^{(3)}$, it is more involved and will be considered elsewhere. The full calculation of $H^{(4)}$ is non-trivial, and we give this expression only to demonstrate what might happen.

The amplitude of the 6 derivative terms may be estimated from Eq.~(\ref{eq49}). For example, we may estimate that 
\be
\overline{{\tilde\partial_k\Phi \tilde\partial^k\Phi\tilde\partial^2\Phi\tilde\partial^2\Phi}}\sim \overline{\tilde\partial_k\Phi \tilde\partial^k\Phi}\;\overline{\tilde\partial^2\Phi\tilde\partial^2\Phi}\sim \left( \frac{3 \Delta_\mathcal{R}}{5
g_{\infty}} \right )^4\left(\frac{k_{eq}}{k_H}\right)^{6}\times 3.9\times F(\kappa_{UV})\,.
\ee
Now, six powers of the equality scale overcome three factors of $\Delta_\mathcal{R}$, using Eq.~(\ref{magic}), making this at least the same size as first-order.
If we take linear theory at face value, and assert that the UV cutoff should be on pc scales, then $\kappa_{UV}\sim 10^8$, and $F\sim 10^5\sim\Delta_\mathcal{R}^{-1}$. Plugging in the numbers, we find these contributions to the Hubble rate~-- very naively~-- are of order unity. That is truly astonishing. This result is very sensitive to the UV cutoff, as well as the baryon fraction which helps reduce small-scale power. 

If we consider what will happen to other quantities at fourth-order, it is clear that such six-derivative-with-four-Phi terms appear also from purely relativistic degrees of freedom. For example, in the Riemann tensor terms such as $(\partial_{\<i} h_{jk\>})^2$ will appear. Using the fact that $h_{ij}\sim\Phi\partial_{\<i}\partial_{j\>}\Phi$, coupled with the fact that time derivatives give factors of $k$ for gravitational waves, we see that such relativistic contributions can also be large; this cannot be associated with any Newtonian effect. Do they add up to something substantial?

Perturbation theory suffers from higher-derivative terms because of the assumed scale-invariant primordial power spectrum, with mild damping of short wavelengths during the radiation era. Without this damping, even the $\Phi^2$ terms would lead to significant backreaction, as modes all the way up to the inflationary cutoff would be included. As it is, contributions for $\Phi^2$ and $\Phi\partial^2\Phi$ are cutoff by the Hubble scale at equality ($k_{eq}$), which is huge in comparison to more fundamental cutoffs. This suppresses modes shorter than the equality scale, but only in a power-law way, not exponentially. Yet it only takes 4 derivatives to overcome this power-law suppression at second-order:  $(\partial^2\Phi)^2$. Exponential suppression of modes below around the Silk scale, or some hypothetical non-linear scale of about that size, would keep such terms just about under control~\cite{Clarkson:2009hr}.


One can argue that we should insert a non-linear cutoff scale $k_{NL}$ into our integrals over $\mathcal{P}_\Phi$ to represent the effective reduction of power expected on small scales which hides the UV divergence. Indeed~\cite{Baumann:2010tm} claim a relativistic virial theorem provides such a scale. Of course, such a theorem cannot exist in general because energy is radiated to infinity in GR, and only the stationary part of a system virialises~\cite{1994CQGra..11..443G}. Nothing dynamic is truly isolated. This is realised in a non-trivial gravitational wave background from structure formation~\cite{Baumann:2007zm}. But this idea does imply that generically, at a fixed time, UV divergences represent something unphysical and can be renormalised away at that time (in~\cite{Baumann:2010tm} they deal with divergences by using a smoothing process on the background, which is effectively a UV cutoff). It is precisely this renormalisation which may lead to a significant backreaction.

In integrals over the linear \emph{primordial} power spectrum, however, seem to be the wrong place to insert a UV cutoff. While one might expect a reduction of power in the \emph{overall} gravitational potential $\Phi^{(\text{total})}$ today (if such a thing were meaningful) compared to $\Phi$, with a change in behaviour at some scale $k_{NL}$ where linear theory becomes inaccurate, \emph{this should come about naturally from higher-order perturbation theory}. If we actually knew the late time power spectrum everything would be fine~-- but an important aim is to calculate it directly from initial conditions, especially when we want to know how the background is renormalised by structure formation. In effect, we do not know what the relation of the background today is to the background at early times (if backreaction turns out to be an important effect).  Higher-order perturbation theory presumably has to rely on the power at small scales in $\Phi$ to calculate $\Phi^{(\text{total})}$. There is no a priori reason to expect small scales to decouple from large order-by-order, even if they actually do decouple physically  for `virialised' regions after the series is summed.  So, an ad hoc insertion into integrals over $\Phi$ doesn't seem to be the solution. It seems higher-order perturbation theory should be considered in detail to solve this problem.

\section{Discussion}

We have discussed here some examples of where the backreaction problem in the standard model of cosmology is non-trivial and not \emph{necessarily} negligible. The two key examples we have discussed are the cases of the deceleration parameter which gives a dimensionless measure of the second time derivative of the scale factor, and the Hubble rate evaluated at fourth-order. In these cases perturbation theory does not give sensible answers because of terms which have significant power in the UV.\footnote{In~\cite{Barausse:2005nf,Kolb:2005me} it was claimed that a divergence in the IR can cause a large backreaction effect, whereas~\cite{Flanagan:2005dk,Hirata:2005ei,Geshnizjani:2005ce} have strongly disagreed with this idea. To make such terms large requires fluctuations out to enormous scales beyond the horizon. The UV divergence, by contrast, arises from the use of scale invariant initial conditions all the way down to small scales, which is usually found in many models of inflation. } 

In some respects dynamical backreaction is small by good fortune: the universe is so hot and had such a long radiation era that small-scale power is significantly reduced over its scale-invariant initial conditions (assuming those to be the case).  The backreaction terms in the Hubble expansion, $\Phi\tilde\partial^2\Phi$, are also the largest ones which appear in the lhs of the Einstein Field Equations, because the Einstein tensor has at most two derivatives of the metric in it. 
In some respects then this settles it: backreaction is small by virtue of there being a very \emph{small} hierarchy of scales between the Hubble scale at equality and the Hubble scale today (they are only a factor of 50 apart in comoving terms). In this evaluation of backreaction, then, what happens on scales smaller than the equality scale is of little relevance. This is perhaps counter-intuitive given how we normally think of backreaction arising from small-scale structure~-- it is really power on very large scales which are responsible for the backreaction effect.

On the other hand, however, we should be able describe the universe using a tetrad version of the Field Equations, or the covariant formulation given in the appendix. This is entirely equivalent to the EFE, but reformulates gravity as a system of first-order PDE's in the Ricci rotation coefficients and Weyl curvature tensor. In such a formulation higher derivative terms appear which are divergent implying backreaction may not be irrelevant. Furthermore, we should also be able to make sense of physical observables such as the fundamental distance-redshift relation, or the geodesic deviation equation.  In these cases, terms such as $(\tilde\partial^2\Phi)^2$ occur frequently, and as we have seen can appear in both their connected and non-connected forms giving rise to divergent backreaction terms, at least on small scales. 

In Buchert's interpretation of backreaction, where $q_\CD$ describes the deceleration of the average scale factor, the divergence is neatly controlled by the domain size, and consequently has an elegant and straightforward interpretation, although it is not observable. This also appears to be robust against various possible gauge effects such as averaging hypersurface~\cite{Umeh:2010pr}. The fact that there is a significant difference between the `cosmological' deceleration and one which is smoothed on scales of a few Mpc, say, is expected since that is the scale where the Hubble flow kicks in. 

On the other hand, our two other definitions of the deceleration parameter~-- which do not depend on Riemannian averaging~-- reveal significant problems. Consider $q_\Theta$ and $q^\text{obs}_0$~-- the local deceleration parameter defined relative to the dust observers, and the observed one defined through the distance-redshift relation. If we calculate the expectation value of either of these we get \emph{enormous} terms. Consider cutting off at a scale suggested by either a scale associated with the end of inflation, or from the dark matter suppression scale, which is around pc scales. Then we have $\kappa_{\text{UV}}\sim 1\,\text{pc}^{-1}/(100\,\text{Mpc})^{-1}\sim 10^8$. Our divergent integral then gives $F(10^8)\sim 10^5$\,! Of course, we have assumed a purely dark matter transfer function, and there is extra suppression from baryons on small scales, but this only reduces it by an order of magnitude for 20\% baryons. One can reformulate the UV cut-off as a smoothing of the first-order potential to give something like the same interpretation one might give to Buchert's $q_\mathcal{D}$, but this is rather ad hoc in this context~-- for $\<q_\Theta\>_\CD$, smoothing has already been done! Instead it maybe tells us that smoothing is necessary order by order in perturbation theory. That is, before constructing second-order perturbation theory, one \emph{necessarily} must smooth structure below a certain scale. But why?  Does this imply that the very notion of ergodicity needs to be made Riemannian: should ensemble averages be replaced by \emph{Riemannian} spatial averages, rather than spatial averages on the background?

This is very unsatisfactory. It is easy to devise universes where the magic number  $\Delta_{\cal R}^2 (k_{eq}/ k_H)^2\sim 1$ (though they may not be anthropically relevant). In an Einstein-de Sitter model, for example, one simply has to wait until such a time as $H_0$ is small enough (though that doesn't happen with LCDM). If there were no radiation era things would be much worse. In such a case perturbation theory would simply break down at second-order and other formalisms for describing structure would have to be devised.  The damping which happens at linear order to suppress a scale-invariant power spectrum would arise from non-linear structure formation, as fully non-linear GR is a well behaved theory.

In our universe, where, apparently coincidentally, $\Delta_{\cal R} (k_{eq}/ k_H)^2\sim1$, it is $\Delta_{\cal R}^2 (k_{eq}/ k_H)^4\sim 1$ which seems to signal problems, because it is accompanied by a factor of $F(\kappa_{UV})$. We can ignore this in many applications of perturbation theory because it simply does not appear. Should we be worried that they do appear in the examples calculated here? Perhaps not. Maybe there are ways around them, or perhaps they are some strange gauge artefact~-- though this seems difficult to realise since the distance-redshift relation is a physical observable (though its ensemble average is not). But we should be worried by the fact that (the ensemble average of) 2nd order perturbations are the same size as first, and 4th might be the same size as second~-- or larger if the $(\partial^2\Phi)^2$ terms are to be taken at face value. It appears as if fourth-order perturbations could give significant changes to the Hubble rate, even if one can dismiss the divergent examples we have discussed at second-order. Generically, we can see why $(\partial^2\Phi)^2$ terms must appear at fourth-order because second-order potential terms are $\Phi\partial^2\Phi$; second-order gradient terms (such as $v_{(2)}^i$) are then $\sim\partial\Phi\partial^2\Phi$, and so have large expectation values when squared. It is tempting to assume that any terms with $(\partial^2\Phi)^2$ in them all must cancel, or must be a gauge effect. However, as we have discussed, in models with $\Phi\partial^2\Phi\sim1$, we do not expect to have large terms cancelling just because they are large, so it need not be the case with $(\partial^2\Phi)^2$ either. It is actually a combination of these terms together with scale invariant initial conditions which causes the problems. If they really do cancel in all possible models, at all orders in perturbation theory, this would be a very persuasive argument that backreaction is indeed small; if not, it is hard to say either way.

Will perturbation theory even converge? Renormalization methods have been devised for Newtonian gravity~\cite{Crocce:2005xy,Crocce:2007dt} which is comforting, but only up to a point: vector modes which represent relativistic frame dragging have no Newtonian counterpart, are not that much smaller than $\Phi^{(2)}$~\cite{Lu:2008ju}; gravitational waves induced by linear scalars are larger and their time derivative is the same size as $\Phi^{(2)}$~\cite{Baumann:2007zm}. These will easily mix together in any proper resummation scheme. However close Newtonian gravity may seem to relativity, GR is the theory of gravity we must work with to address these issues.



\ack

We thank Ruth Durrer, George Ellis, Julien Larena, Roy Maartens and Jean-Philippe Uzan for discussions.
This work is supported by the NRF (South Africa) and the SKA (South Africa).


\appendix

\section{covariant formulation of the field equations}

The 1+3 covariant Ehlers-Ellis formalism provides a physically transparent formulation of the field equations full nonlinear generality (see~\cite{Ellis:1998ct,Tsagas:2007yx} for reviews). The Ehlers-Ellis formalism is a covariant Lagrangian approach to gravitational dynamics, based on a decomposition relative to a chosen 4-velocity field $u^\a$.
The fundamental tensors are
\begin{equation}\label{hep}
h_{\a\b}=g_{\a\b}+u_\a u_\b,~ ~ \ep_{\a\b c}=\eta_{\a\b c d}u^ d,
\end{equation}
where $h_{\a\b}$ projects into the instantaneous rest space of comoving
observers, and $\ep_{\a\b c}$ is the projection of the spacetime alternating tensor $\eta_{\a\b c  d}=-\sqrt{-g}
\delta^0{}_{[\a}\delta^1{}_\b\delta^2{}_ c\delta^3{}_{ d]}$, and so
\begin{equation}
\eta_{abcd} = 2u_{[a}\ep_{b]cd}-2\ep_{ab[c}u_{d]}\,,~
\ep_{abc}\ep^{def}=3!h_{[a}{}^dh_b{}^eh_{c]}{}^f\,.
\end{equation}
The projected symmetric tracefree (PSTF) parts of vectors and
rank-2 tensors are
\begin{eqnarray}
V_{\langle \a\rangle}=h_\a{}^\b V_\b\,,~ S_{\langle \a\b \rangle }=
\Big\{h_{(a}{}^ c h_{\b)}{}^ d-
{{1\over3}}h^{ c  d }h_{\a\b }\Big\}S_{ c  d }\,. \label{pstf}
\end{eqnarray}
The skew part of a
projected rank-2 tensor is spatially dual to the projected vector,
$S_{\a}={1\over2}\ep_{\a\b  c}S^{[\b  c]}$, and then any projected rank-2
tensor has the decomposition
$S_{\a\b }={1\over 3}Sh_{\a\b }+\ep_{\a\b  c}S^{ c}+S_{\langle \a\b \rangle}$, where $S=S_{ c d} h^{ c  d}$.

The covariant derivative $\nabla_{\a}$ defines 1+3 covariant time and spatial derivatives:
\begin{eqnarray}
\dot{J}^{\a\cdots}{}{}_{\cdots \b}= u^{ c} \nabla_{ c}
J^{\a\cdots}{}{}_{\cdots \b},\, \D_{ c} J^{\a\cdots}{}{}_{\cdots \b} =    h_{ c}{}^ d h^{\a}{}_ e\cdots h_{\b}{}^f
\nabla_ d J^{ e\cdots}{}{}_{\cdots f}. \label{dd}
\end{eqnarray}
The projected derivative $\D_{\a}$
defines a covariant PSTF divergence, $\div V=\D^\a V_\a\,, ~ \div S_a=\D^\b S_{\a\b}$,
and a covariant PSTF curl,
\begin{eqnarray}
\c V_{\a}=\ep_{\a\b c}\D^{\b}V^{ c}\,,~ \c
S_{\a\b }=\ep_{ c  d (\a}\D^{ c}S_{\b)}{}^ d\,. \label{curl}
\end{eqnarray}

The relative motion of comoving observers
is encoded in the PSTF kinematical quantities: the volume expansion rate,  4-acceleration, vorticity
and shear, given respectively by
\begin{eqnarray}
\Theta=\D^{\a}u_{\a},~ A_{\a}=\dot{u}_{\a},~ {\omega}_\a=\c u_\a ,~ \sigma_{\a\b }=\D_{\langle \a}u_{\b\rangle }. \label{kin}
\end{eqnarray}
Thus
\begin{eqnarray}
\nabla_{\b}u_{\a}={{1\over3}}\Theta h_{\a\b }+\ep_{\a\b  c}\omega^{ c}
+\sigma_{\a\b }-A_{\a}u_{\b}\,. \label{du}
\end{eqnarray}

The PSTF dynamical quantities describe the
sources of the gravitational field:
the (total) energy density $\rho=T_{\a\b }u^{\a}u^{\b}$, isotropic pressure
$p={1\over3}h_{\a\b }T^{\a\b }$, momentum density $q_{\a}=-T_{\langle \a\rangle \b}u^{\b}$,
and anisotropic stress $\pi_{\a\b }=T_{\langle \a\b \rangle}$, where $T_{\a\b }$ is
the total energy-momentum tensor. The locally free gravitational
field, i.e. the part of the spacetime curvature not directly
determined locally by dynamical sources, is given by the Weyl tensor
$C_{\a\b  c  d }$. This splits into the PSTF gravito-electric and gravito-magnetic fields
\begin{eqnarray}
E_{\a\b }=C_{\a  c \b  d}u^{ c}u^ d\,,~~
H_{\a\b }={{1\over2}}\ep_{\a c  d }C^{ c  d }{}{}_{\b  e}u^ e  \,,
\end{eqnarray}
which provide a covariant description of tidal forces
and gravitational radiation.

The Ricci and Bianchi identities,
\begin{eqnarray}
\nabla_{[\a} \nabla_{\b]} u_ c= R_{\a\b  c   d}u^{ d}, ~~~ \nabla^ d
C_{\a\b  c  d } =- \nabla_{[\a}\Big\{ R_{\b] c} - {1\over6}Rg_{\b] c} \Big\}, \label{rbi}
\end{eqnarray}
produce the
fundamental evolution and constraint equations governing the
covariant quantities. Einstein's equations are
incorporated via the algebraic replacement of the Ricci tensor
\begin{equation}\label{efe}
R^{\a\b }=T^{\a\b }-{1\over2}T_{ c}{}^{ c}g^{\a\b }+\Lambda g^{\a\b},
\end{equation}
where $T^{\a\b}$ is the total energy-momentum tensor.

The resulting equations, in fully nonlinear form and for a general
source of the gravitational field, are:\\
\noindent{\em Evolution:}
\begin{eqnarray}\fl
 \dot{\rho} +(\rho+p)\Theta+\div q = -2A^{\a} q_{\a}
 -\sigma^{\a\b }\pi_{\a\b }\,, 
 \label{e1}\\ \fl
 \dot{\Theta} +{{1\over3}}\Theta^2 +{{1\over2}}(\rho+3p)-\Lambda-\div
A 
= -\sigma_{\a\b }\sigma^{\a\b }
+2\omega_{\a}\omega^{\a}+A_{\a}A^{\a} \,,
\label{e2}\\ \fl
 \dot{q}_{\langle \a\rangle }
+{{4\over3}}\Theta q_{\a}+(\rho+p)A_{\a} +\D_{\a} p +\div\pi_{\a} 
= -\sigma_{\a\b }q^{\b}
+\ep_{\a\b c}\omega^\b q^{ c} -A^{\b}\pi_{\a\b } \,,
\label{e3} \\ \fl
 \dot{\omega}_{\langle \a\rangle } +{{2\over3}}\Theta\omega_{\a}
+{{1\over2}}\c A_{\a} = \sigma_{\a\b }\omega^{\b} \,,\label{e4}
\\ \fl
 \dot{\sigma}_{\langle \a\b \rangle } +{{2\over3}}\Theta\sigma_{\a\b }
+E_{\a\b }-{{1\over2}}\pi_{\a\b } -\D_{\langle \a}A_{\b\rangle } 
= -\sigma_{ c\langle \a}\sigma_{\b\rangle }{}^{ c} - \omega_{\langle \a}\omega_{\b\rangle }
+A_{\langle \a}A_{\b\rangle }\,,
\label{e5}\\ \fl
 \dot{E}_{\langle \a\b \rangle } +\Theta E_{\a\b }
-\c H_{\a\b } +{{1\over2}}(\rho+p)\sigma_{\a\b }
+{{1\over2}}
\dot{\pi}_{\langle \a\b \rangle } +{{1\over6}}
\Theta\pi_{\a\b } +{{1\over2}}\D_{\langle \a}q_{\b\rangle } 
\\ \fl~
=-A_{\langle \a}q_{\b\rangle } +2A^{ c}\ep_{ c  d (\a}H_{\b)}{}^ d
+3\sigma_{ c\langle \a}E_{\b\rangle }{}^{ c} \nonumber
-\omega^{ c} \ep_{ c  d (\a}E_{\b)}{}^ d -{{1\over2}}\sigma^{ c}{}_{\langle
\a}\pi_{\b\rangle  c} -{{1\over2}}\omega^{ c}\ep_{ c  d (\a}\pi_{\b)}{}^ d \,,
\label{e6}\\ \fl
 \dot{H}_{\langle \a\b \rangle } +\Theta H_{\a\b } +\c E_{\a\b }
-{{1\over2}}\c\pi_{\a\b } 
= 3\sigma_{ c\langle \a}H_{\b\rangle }{}^{ c}
-\omega^{ c} \ep_{ c  d (\a}H_{\b)}{}^ d \nonumber\\
~~~~~~~~~ -2A^{ c}\ep_{ c  d (\a}E_{\b)}{}^ c -{{3\over2}}\omega_{\langle \a}q_{\b\rangle
}+ {{1\over2}}\sigma^{ c}{}_{(\a}\ep_{\b) c  d }q^ d \,. \label{e7}
\end{eqnarray}

\noindent{\em Constraint:}
\begin{eqnarray}
\fl
 \div\omega = A^{\a}\omega_{\a} \,, ~~~~~~~~~~~~~~~~~~~~~~~~~~~~~ ~~~~~~~~~~~~~~~~~~~~~~~~~~~~~~~~~~~~~~~~~ \label{c1}\\ \fl
 \div\sigma_{\a}-\c\omega_{\a} -{{2\over3}}\D_{\a}\Theta +q_{\a} =-
2\ep_{\a\b c}\omega^\b A^{ c}  \,,\label{c2}\\ \fl
  \c\sigma_{\a\b }+\D_{\langle \a}\omega_{\b\rangle }
 -H_{\a\b }= -2A_{\langle \a}
\omega_{\b\rangle } \,,\label{c3}\\\fl
\div E_{\a}
+{{1\over2}}\div\pi_{\a}
 -{{1\over3}}\D_{\a}\rho
+{{1\over3}}\Theta q_{\a} \nonumber\\\fl
 ~~~~~~= \ep_{\a\b c}\sigma^\b{}_ d H^{ c  d} -3H_{\a\b}
\omega^{\b} +{{1\over2}}\sigma_{\a\b }q^{\b}-{{3\over2}}
\ep_{\a\b c}\omega^\b q^{ c}  \,,\label{c4}\\\fl
 \div H_{\a}
+{{1\over2}}\c q_{\a}
 -(\rho+p)\omega_{\a}\nonumber\\\fl
 ~~~~~~ =
-\ep_{\a\b c}\sigma^\b{}_ d E^{ c  d}-{{1\over2}}\ep_{\a\b c}\sigma^\b{}_ d \pi^{ c  d}   +3E_{\a\b }\omega^{\b}
-{{1\over2}}\pi_{\a\b } \omega^{\b}  \,.\label{c5}
\end{eqnarray}
The energy and momentum conservation equations are the evolution equations~(\ref{e1}) and (\ref{e3}). The dynamical quantities $\rho, p, q_\a, \pi_{\a\b}$ in the evolution and constraint equations
(\ref{e1})--(\ref{c5}) are the total quantities, with
contributions from all dynamically significant particle species.

\subsection{Moment Decomposition}

Here we give a summary of the covariant spherical harmonics we use, following~\cite{2000AnPhy.282..285G}.

An observer moving with $4-$velocity $u^a$ at position $x^i$, in a direction $e^a$ on the unit sphere
$ \left( e^ae_a=1, e^au_a \right)$ measures the luminosity of a distant supernova or galaxy. The direction $e^a$ can be given in terms of an orthonormal tetrad frame as, for example 
\begin{equation}
e^a(\theta, \phi)=\left(0,\sin \theta\sin\phi,\sin\theta\cos\phi, \cos\theta\right)
\end{equation}
There is a $1-1$ mapping between all symmetric trace-free tensors of rank $l$ and the spherical harmonics of order $\ell$~\cite{Pirani,1981MNRAS.194..439T,2000AnPhy.282..285G}. 
Consider a spherical harmonic decomposition of $f(x^i;\theta,\phi)$,
\begin{equation}
f=\sum_{l=0}^\infty f_l=\sum_{\ell=0}^{\infty}\sum_{m=-\ell}^{+\ell} F_\ell^m(x^i)Y_\ell^m(\theta ,\phi )
\label{Sharmonic}
\end{equation}
where $Y_\ell^m(\theta ,\phi )$ are the usual surface spherical harmonic.
An alternative way to carry out this harmonic  expansion is to write $f$ in the form
\begin{equation}
f=\sum_{\ell=0}^{\infty} F_{A_\ell}{e}^{A_\ell} = F+ F_ae^a + F_{ab}e^ae^b + F_{abc}e^ae^be^c+ F_{abcd}e^ae^be^ce^d+....
\end{equation}
where the spherical harmonic coefficients $F_{A_\ell}$ are symmetric, trace-free tensors orthogonal to $u^a$:
\begin{eqnarray}
F_{A_{\ell}}=F_{(A_{\ell})}\,,~~F_{A_{\ell} ab} h^{ab}
=0\,,~~F_{A_{\ell} a} u^a \,.\label{symmetry}
\end{eqnarray}
Round brackets ``(..)'' denote the symmetric part of a set of
indices, angle brackets ``$\la..\ra$'' the PSTF part of the indices. We use the shorthand notation using the compound index 
$A_{\ell} = a_1 a_2 ... a_{\ell}$, and ${e}^{A_\ell}=e^{a_1}\cdots e^{a_\ell}$. 
The PSTF part of $e^{\la A_\ell \ra}$ \cite
{Pirani,Thorne:1980ru,1981MNRAS.194..439T,1983AnPhy.150..455E} is
\begin{eqnarray}
e^{\la A_\ell \ra}=\sum_{k=0}^{[\ell /2]}
B_{\ell k}h^{(a_1a_2}\cdots h^{a_{2k-1}a_{2k}}e^{a_{2k+1}}e^{a_{2k+2}}\cdots e^{a_\ell)}\,, 
\label{II-3}
\end{eqnarray}
where $B_{\ell k}$ are given by~\cite{Pirani}, 
\begin{eqnarray}
B_{\ell n}={\frac{{(-1)^n\ell !(2\ell -2n-1)!!}}{{(\ell -2n)!(2\ell
-1)!!(2n)!!}}}.  \label{B} 
\end{eqnarray}
Here $[\ell /2]$ means the largest integer part less than or equal to
$\ell /2$, and $\ell !! =\ell (\ell
-2)(\ell -4)\cdots(2~~or~~1)$. The normalization for $e^{A_{\ell}}$ is given by \cite{1983AnPhy.150..455E}, for
 odd and even $\ell$ respectively: 
\begin{eqnarray}
\frac{1}{4 \pi} \int_{4 \pi} e^{A_{2 \ell+1}} d \Omega = 0,
~~\mbox{and}~~  
\frac{1}{4 \pi} \int_{4 \pi} e^{A_{2 \ell}} d \Omega = \frac{1}{2 \ell+1}
h^{(A_{2 \ell})}\,,
\end{eqnarray}
which implies
\begin{eqnarray}
\int_{4\pi }e^{A_\ell}e^{B_m}d\Omega ={\frac{4\pi
}{\ell+m+1}}h^{(A_\ell B_m)}.
\label{orth1}
\end{eqnarray}
if $\ell+m$ is even, and is zero otherwise.
The orthogonality condition for ${e}^{\<A_\ell\>}$ then follows:
\begin{eqnarray}\fl
\int d\Omega e^{\<A_\ell\> }e_{\<B_m\>}=\delta _m^l\Delta _\ell h^{\la a_1}{}_{\la
b_1}\cdots h^{a_\ell \ra }{}_{b_\ell \ra }~~\mbox{with}~~
\Delta _\ell ={\frac{4\pi }{(2\ell +1)}}{\frac{2^\ell (\ell !)^2}
{(2\ell )!}}\,,  \label{Delta-ortho0-01}
\end{eqnarray}
which then implies the relation between $f$ and its spherical harmonic moments $f_{A_\ell}$:
\be
F _{A_\ell }(x^i)=\Delta _\ell ^{-1}\int_{4\pi }d\Omega 
\,e_{\<A_\ell\> }\,f(x^i,e^a)\,.
\ee
Note the recursion relation
\begin{eqnarray}
e^{\<A_{\ell +1}\>}=e^{(a_{\ell +1}}e^{\<A_\ell\> )}-{\frac{\ell ^2}{(2\ell
+1)(2\ell -1)}} h^{(a_{\ell +1}a_\ell }e^{\<A_{\ell -1}\>)}  \label{PSTF_rec}
\end{eqnarray}
relates the $(\ell +1)-th$ term to the $\ell -th$ term and the $(\ell
-1)-th$ term.

\section{The deceleration parameters}

We give here the full deceleration parameters for the cases we consider. First the deceleration parameter associated with the fluid expansion rate $\Theta$:
\begin{eqnarray}\fl
q_{\Theta} =   -1+ \frac{3 }{2}\Omega_m - 3(1+\hat{g})(1-\Omega_m)\Phi+ \left[(1+\sfrac{2}{3}\hat{g})-\frac{4}{9\Omega_m}(1+\hat{g})\right]\tilde{\partial}^2\Phi  \nonumber \\ \nonumber  
-3\hat{g}\left(4+\sfrac{3}{2}\hat{g}\right)(1-\Omega_m)\,\Phi^2-\frac{3}{2H}(1-\Omega_m)(H\Phi^{(2)}+\dot{\Psi}^{(2)})
\\ \nonumber
+\frac{a}{6H}(1-3\Omega_m)\left(H\tilde{\partial}_kv_{(2)}^k-\tilde{\partial}_k\dot{v}_{(2)}^k\right) -\frac{1}{6}\left(\tilde{\partial}^2\Phi^{(2)}-\tilde{\partial}^2\Psi^{(2)}\right)
\\ \nonumber  
+ \frac{1}{9\Omega_m}\left[3\Omega_m(11+14\hat{g}+6\hat{g}^2)-(16+36\hat{g}+2\hat{g}^2)\right]\Phi\tilde{\partial}^2\Phi \\ \nonumber  
-\frac{1}{27\Omega^2_m}\left[\sfrac{27}{6}\Omega^2_m(11+12\hat{g})-24\Omega(1-\hat{g})-4(1+2\hat{g}+\hat{g}^2)\right] \tilde{\partial}_k\Phi\tilde{\partial}^k\Phi\\ \nonumber  
-\frac{4}{27\Omega^2_m}(1+2\hat{g}+\hat{g}^2)\tilde{\partial}^k\Phi\tilde{\partial}_k\tilde{\partial}^2\Phi \\ \nonumber  
+\frac{2}{27 \Omega_m}\left[\Omega_m(5+8\hat{g}+3\hat{g}^2)-\sfrac{8}{3}(1+2\hat{g}+\hat{g}^2)\right]\tilde{\partial}^2\Phi\tilde{\partial}^2\Phi\,.
\end{eqnarray}
The deceleration parameter an observer would measure from the all-sky average of the redshift-distance relation:
\begin{eqnarray}\fl
\mathcal{Q} =  -1+ \frac{3 }{2}\Omega_m - 3(1+\hat{g})(1-\Omega_m)\Phi+ \left[(1+\sfrac{2}{3}\hat{g})-\frac{4}{9\Omega_m}(1+\hat{g})\right]\tilde{\partial}^2\Phi \nonumber 
\\ \nonumber  
-3\hat{g}\left(4+\sfrac{3}{2}\hat{g}\right)(1-\Omega_m)\,\Phi^2
-\frac{3}{2H}(1-\Omega_m)(H\Phi^{(2)}+\dot{\Psi}^{(2)})
\\ \nonumber  
- \sfrac{1}{2}{a\Omega_m\tilde{\partial}_kv_{(2)}^k}- \frac{a}{3H}{\tilde{\partial}_k\dot{v}_{(2)}^k} 
-\sfrac{1}{3}\tilde{\partial}^2\Phi^{(2)}+\sfrac{1}{6}\tilde{\partial}^2\Psi^{(2)}
\\ \nonumber  
+ \frac{1}{9\Omega_m}\left[3\Omega_m(10+14\hat{g}+6\hat{g}^2)-8(2+5\hat{g}+3\hat{g}^2)\right]\Phi\tilde{\partial}^2\Phi 
\\ \nonumber  
-\frac{1}{\Omega^2_m}\left[\sfrac{1}{6}\Omega_m^2(13+12\hat{g})-\sfrac{4}{9}\Omega_m(2-\hat{g}-12\hat{g}^2) -\sfrac{4}{27}(1+2\hat{g}+\hat{g}^2)\right]\tilde{\partial}_k\Phi\tilde{\partial}^k\Phi
\\ \nonumber  
+\frac{4}{27\Omega_m^2}(1+2\hat{g}+\hat{g}^2)\tilde{\partial}_i\tilde{\partial}_j\Phi\tilde{\partial}^i\tilde{\partial}^j\Phi
-\frac{8}{27\Omega_m^2}(1+2\hat{g}+\hat{g}^2)\tilde{\partial}^k\Phi\tilde{\partial}^2\tilde{\partial}_k\Phi
\\ \nonumber  
+\frac{2}{27 \Omega_m}\left[\Omega_m(5+8\hat{g}+3\hat{g}^2)-4(1+2\hat{g}+\hat{g}^2)\right]\tilde{\partial}^2\Phi\tilde{\partial}^2\Phi
\end{eqnarray}

\noindent Finally, we give the deceleration parameter defined via Eq.~(\ref{qDdef}), averaged in the local rest-frame of the dust observers:
\begin{eqnarray}\fl
q_{\mathcal{D}} =   -1+ \frac{3}{2}\Omega_m - 3\left[(1+\hat{g})(1-\Omega_m)+\sfrac{3}{2}\Omega_m g_IH\right]\<\Phi\>
+ 
\left[1+\sfrac{2}{3}\hat{g}-\frac{4}{9\Omega_m}(1+\hat{g})\right]\<\tilde{\partial}^2\Phi \>\nonumber\\ \nonumber  
+\left\{3\left(4+\hat{g}-\sfrac{1}{2}\hat{g}^2\right)+9g_IH\left[(1+\hat{g})(1-\Omega_m)+\sfrac{3}{4}g_IH\right]-\sfrac{27}{2}g_I^2H^2\Omega_m^2\left(\sfrac{5}{4}-\Omega_m\right)\right\}\<\Phi^2\>\\ \nonumber  
-3\left\{(4+5\hat{g}) +\sfrac{3}{2}\Omega_m(3+4\hat{g}+\hat{g}^2)-3g_IH\left[(1+\hat{g})\left(1-\sfrac{1}{2}{\Omega_m^2}\right)-\Omega_m(1+2\hat{g}+4g_IH)\right]\right.\\ \nonumber  \left.
-\sfrac{27}{2}\Omega_m^2g_I^2H^2\left(1-\sfrac{1}{2}{\Omega_m}\right)\right\}\<\Phi\>^2\\ \nonumber  
-\frac{1}{\Omega_m^2}\left[\sfrac{1}{2}{\Omega_m^2}(3+4\hat{g})-\sfrac{4}{9}\Omega_m(2+\hat{g}-\hat{g}^2)-\sfrac{4}{27}(1+2\hat{g}+\hat{g}^2)\right]\<\tilde{\partial}_k\Phi\tilde{\partial}^k\Phi\>\\ \nonumber  
+\frac{1}{3}\left\{(17+20\hat{g}+6\hat{g}^2)-\frac{2}{3\Omega_m}(10+14\hat{g}+4\hat{g}^2)\right.\\ \nonumber  \left.
+g_IH\left[(1-2\hat{g})-\frac{4}{3\Omega_m}(1+2\hat{g})-\Omega_m(2+\hat{g})\right]\right\}\<\Phi\>\<\tilde{\partial}^2\Phi\>\\ \nonumber  
-\frac{3}{2H}(1-\Omega_m)\left(H\<\Phi^{(2)}\>+\<\dot\Psi^{(2)}\>\right)
+\sfrac{1}{6}a(2-3\Omega_m)\<\tilde{\partial}_kv_2^k\>+\sfrac{1}{6}(\<\tilde{\partial}^2\Phi^{(2)}\>\\ \nonumber  
+\frac{9}{8}\Omega_mH
\int^t\d t' \left[\<\Phi^{(2)}\>- \sfrac{1}{2} \<\Phi^2\> - \<v_1^k v_{1k} \>
  - \frac{g_IH}{a}\<v_1^k\tilde{\partial}_k \Phi\>\right]\\  
-\frac{2}{27\Omega_m}\left[\sfrac{2}{3}(1+2\hat{g}+\hat{g}^2)-\Omega_m(5+8\hat{g}+3\hat{g}^3)\right]\<\tilde{\partial}^2\Phi\>^2\,.
\end{eqnarray}
Note that there are no connected $(\partial^2\Phi)^2$ terms in this expression.

\section*{References}

\begin{thebibliography}{10}
\expandafter\ifx\csname url\endcsname\relax
  \def\url#1{{\tt #1}}\fi
\expandafter\ifx\csname urlprefix\endcsname\relax\def\urlprefix{URL }\fi
\providecommand{\eprint}[2][]{\url{#2}}

\bibitem{Ell84}
{Ellis} G~F~R 1984 {Relativistic cosmology - Its nature, aims and problems}
  {\em General Relativity and Gravitation Conference\/} ed {B~Bertotti, F~de
  Felice, \& A~Pascolini} pp 215--288

\bibitem{Buchert:1995fz}
Buchert T and Ehlers J 1997 {\em Astron.Astrophys.\/} {\bf 320} 1--7
  (\textit{Preprint} \eprint{astro-ph/9510056})

\bibitem{Buchert:1999er}
Buchert T 2000 {\em Gen. Rel. Grav.\/} {\bf 32} 105--125 (\textit{Preprint}
  \eprint{gr-qc/9906015})

\bibitem{Buchert:2001sa}
Buchert T 2001 {\em Gen. Rel. Grav.\/} {\bf 33} 1381--1405 (\textit{Preprint}
  \eprint{gr-qc/0102049})

\bibitem{Rasanen:2003fy}
Rasanen S 2004 {\em JCAP\/} {\bf 0402} 003 (\textit{Preprint}
  \eprint{astro-ph/0311257})

\bibitem{Barausse:2005nf}
Barausse E, Matarrese S and Riotto A 2005 {\em Phys. Rev.\/} {\bf D71} 063537
  (\textit{Preprint} \eprint{astro-ph/0501152})

\bibitem{Kolb:2005da}
Kolb E~W, Matarrese S and Riotto A 2006 {\em New J. Phys.\/} {\bf 8} 322
  (\textit{Preprint} \eprint{astro-ph/0506534})

\bibitem{Rasanen:2006kp}
Rasanen S 2006 {\em JCAP\/} {\bf 0611} 003 (\textit{Preprint}
  \eprint{astro-ph/0607626})

\bibitem{Kasai:2007fn}
Kasai M 2007 {\em Prog. Theor. Phys.\/} {\bf 117} 1067--1075 (\textit{Preprint}
  \eprint{astro-ph/0703298})

\bibitem{Russ:1996km}
Russ H, Soffel M~H, Kasai M and Borner G 1997 {\em Phys. Rev.\/} {\bf D56}
  2044--2050 (\textit{Preprint} \eprint{astro-ph/9612218})

\bibitem{Kolb:2004am}
Kolb E~W, Matarrese S, Notari A and Riotto A 2005 {\em Phys. Rev.\/} {\bf D71}
  023524 (\textit{Preprint} \eprint{hep-ph/0409038})

\bibitem{Li:2007ci}
Li N and Schwarz D~J 2007 {\em Phys. Rev.\/} {\bf D76} 083011
  (\textit{Preprint} \eprint{gr-qc/0702043})

\bibitem{Li:2007ny}
Li N and Schwarz D~J 2008 {\em Phys. Rev.\/} {\bf D78} 083531
  (\textit{Preprint} \eprint{0710.5073})

\bibitem{Li:2008yj}
Li N, Seikel M and Schwarz D~J 2008 {\em Fortsch. Phys.\/} {\bf 56} 465--474
  (\textit{Preprint} \eprint{0801.3420})

\bibitem{Clarkson:2009hr}
Clarkson C, Ananda K and Larena J 2009 {\em Phys. Rev.\/} {\bf D80} 083525
  (\textit{Preprint} \eprint{0907.3377})

\bibitem{Clarkson:2009jq}
Clarkson C 2010 {\em AIP Conf.Proc.\/} {\bf 1241} 784--796 (\textit{Preprint}
  \eprint{0911.2601})

\bibitem{Chung:2010xx}
{Chung} H 2010 {\em ArXiv e-prints\/} (\textit{Preprint} \eprint{1009.1333})

\bibitem{Umeh:2010pr}
{Umeh} O, {Larena} J and {Clarkson} C 2010 {\em ArXiv e-prints\/}
  (\textit{Preprint} \eprint{1011.3959})

\bibitem{EllSto87}
{Ellis} G~F~R and {Stoeger} W 1987 {\em Class. Quant. Grav.\/} {\bf 4}
  1697--1729

\bibitem{Ishibashi:2005sj}
Ishibashi A and Wald R~M 2006 {\em Class. Quant. Grav.\/} {\bf 23} 235--250
  (\textit{Preprint} \eprint{gr-qc/0509108})

\bibitem{Flanagan:2005dk}
Flanagan E~E 2005 {\em Phys. Rev.\/} {\bf D71} 103521 (\textit{Preprint}
  \eprint{hep-th/0503202})

\bibitem{Hirata:2005ei}
Hirata C~M and Seljak U 2005 {\em Phys. Rev.\/} {\bf D72} 083501
  (\textit{Preprint} \eprint{astro-ph/0503582})

\bibitem{Geshnizjani:2005ce}
Geshnizjani G, Chung D~J~H and Afshordi N 2005 {\em Phys. Rev.\/} {\bf D72}
  023517 (\textit{Preprint} \eprint{astro-ph/0503553})

\bibitem{Bonvin:2005ps}
Bonvin C, Durrer R and Gasparini M 2006 {\em Phys.Rev.\/} {\bf D73} 023523
  (\textit{Preprint} \eprint{astro-ph/0511183})

\bibitem{Siegel:2005xu}
Siegel E~R and Fry J~N 2005 {\em Astrophys.J.\/} {\bf 628} L1--L4
  (\textit{Preprint} \eprint{astro-ph/0504421})

\bibitem{Giovannini:2005sy}
Giovannini M 2006 {\em Phys.Lett.\/} {\bf B634} 1--4 (\textit{Preprint}
  \eprint{hep-th/0505222})

\bibitem{Bonvin:2006en}
Bonvin C, Durrer R and Kunz M 2006 {\em Phys.Rev.Lett.\/} {\bf 96} 191302
  (\textit{Preprint} \eprint{astro-ph/0603240})

\bibitem{Vanderveld:2007cq}
Vanderveld R~A, Flanagan E~E and Wasserman I 2007 {\em Phys. Rev.\/} {\bf D76}
  083504 (\textit{Preprint} \eprint{0706.1931})

\bibitem{Behrend:2007mf}
Behrend J, Brown I~A and Robbers G 2008 {\em JCAP\/} {\bf 0801} 013
  (\textit{Preprint} \eprint{0710.4964})

\bibitem{Kumar:2008uk}
Kumar N and Flanagan E~E 2008 {\em Phys. Rev.\/} {\bf D78} 063537
  (\textit{Preprint} \eprint{0808.1043})

\bibitem{Rosenthal:2008ic}
Rosenthal E and Flanagan E~E 2008  (\textit{Preprint} \eprint{0809.2107})

\bibitem{Krasinski:2009qq}
Krasinski A, Hellaby C, Celerier M~N and Bolejko K 2010 {\em Gen.Rel.Grav.\/}
  {\bf 42} 2453--2475 (\textit{Preprint} \eprint{0903.4070})

\bibitem{Ziaeepour:2009zu}
{Ziaeepour} H 2009 {\em ArXiv e-prints\/} (\textit{Preprint}
  \eprint{0906.4278})

\bibitem{Tomita:2009ar}
{Tomita} K 2009 {\em ArXiv e-prints\/} (\textit{Preprint} \eprint{0906.1325})

\bibitem{Baumann:2010tm}
{Baumann} D, {Nicolis} A, {Senatore} L and {Zaldarriaga} M 2010 {\em ArXiv
  e-prints\/} (\textit{Preprint} \eprint{1004.2488})

\bibitem{Green:2010qy}
{Green} S~R and {Wald} R~M 2010 {\em ArXiv e-prints\/} (\textit{Preprint}
  \eprint{1011.4920})

\bibitem{Notari:2005xk}
Notari A 2006 {\em Mod. Phys. Lett.\/} {\bf A21} 2997--3001 (\textit{Preprint}
  \eprint{astro-ph/0503715})

\bibitem{Rasanen:2010wz}
Rasanen S 2010 {\em Phys. Rev.\/} {\bf D81} 103512 (\textit{Preprint}
  \eprint{1002.4779})

\bibitem{Rasanen:2009uw}
Rasanen S 2010 {\em JCAP\/} {\bf 1003} 018 (\textit{Preprint}
  \eprint{0912.3370})

\bibitem{1966ApJ...143..379K}
{Kristian} J and {Sachs} R~K 1966 {\em \apj\/} {\bf 143} 379--+

\bibitem{1969CMaPh..12..108E}
{Ellis} G~F~R and {MacCallum} M~A~H

\bibitem{1970CMaPh..19...31M}
{MacCallum} M~A~H and {Ellis} G~F~R 1970 {\em Communications in Mathematical
  Physics\/} {\bf 19} 31--64

\bibitem{Clarksonthesis}
Clarkson C~A 2000 {\em ArXiv Astrophysics e-prints\/} (\textit{Preprint}
  \eprint{arXiv:astro-ph/0008089})

\bibitem{1995ApJ...453..574Z}
{Zotov} N~V and {Stoeger} W~R 1995 {\em \apj\/} {\bf 453} 574--+

\bibitem{Boersma:1997yt}
Boersma J~P 1998 {\em Phys.Rev.\/} {\bf D57} 798--810 (\textit{Preprint}
  \eprint{gr-qc/9711057})

\bibitem{Kolb:2009rp}
Kolb E~W, Marra V and Matarrese S 2010 {\em Gen.Rel.Grav.\/} {\bf 42}
  1399--1412 (\textit{Preprint} \eprint{0901.4566})

\bibitem{Noonan1984}
{Noonan} T~W 1984 {\em General Relativity and Gravitation\/} {\bf 16}
  1103--1118

\bibitem{Noonan1985}
{Noonan} T~W 1985 {\em General Relativity and Gravitation\/} {\bf 17} 535--544

\bibitem{Paranjape:2006ww}
Paranjape A and Singh T~P 2008 {\em Gen. Rel. Grav.\/} {\bf 40} 139--157
  (\textit{Preprint} \eprint{astro-ph/0609481})

\bibitem{Paranjape:2007wr}
Paranjape A and Singh T~P 2007 {\em Phys. Rev.\/} {\bf D76} 044006
  (\textit{Preprint} \eprint{gr-qc/0703106})

\bibitem{Paranjape:2007uj}
Paranjape A 2008 {\em Int. J. Mod. Phys.\/} {\bf D17} 597--601
  (\textit{Preprint} \eprint{0705.2380})

\bibitem{Brown:2008ra}
Brown I~A, Robbers G and Behrend J 2009 {\em JCAP\/} {\bf 0904} 016
  (\textit{Preprint} \eprint{0811.4495})

\bibitem{Stoeger:1999ig}
Stoeger William~R S, Helmi A and Torres D~F 2007 {\em Int.J.Mod.Phys.\/} {\bf
  D16} 1001--1026 (\textit{Preprint} \eprint{gr-qc/9904020})

\bibitem{Eisenstein:1997ik}
Eisenstein D~J and Hu W 1998 {\em Astrophys.J.\/} {\bf 496} 605
  (\textit{Preprint} \eprint{astro-ph/9709112})

\bibitem{Bartolo:2005kv}
Bartolo N, Matarrese S and Riotto A 2006 {\em JCAP\/} {\bf 0605} 010
  (\textit{Preprint} \eprint{astro-ph/0512481})

\bibitem{Mollerach:2003nq}
Mollerach S, Harari D and Matarrese S 2004 {\em Phys. Rev.\/} {\bf D69} 063002
  (\textit{Preprint} \eprint{astro-ph/0310711})

\bibitem{Ananda:2006af}
Ananda K~N, Clarkson C and Wands D 2007 {\em Phys.Rev.\/} {\bf D75} 123518
  (\textit{Preprint} \eprint{gr-qc/0612013})

\bibitem{Baumann:2007zm}
Baumann D, Steinhardt P~J, Takahashi K and Ichiki K 2007 {\em Phys. Rev.\/}
  {\bf D76} 084019 (\textit{Preprint} \eprint{hep-th/0703290})

\bibitem{Lu:2008ju}
Lu T~H~C, Ananda K, Clarkson C and Maartens R 2009 {\em JCAP\/} {\bf 0902} 023
  (\textit{Preprint} \eprint{0812.1349})

\bibitem{Lyth:2007jh}
Lyth D~H 2007 {\em JCAP\/} {\bf 0712} 016 (\textit{Preprint}
  \eprint{0707.0361})

\bibitem{Abramo:2010gk}
Abramo L and Pereira T~S 2010  * Temporary entry * (\textit{Preprint}
  \eprint{1002.3173})

\bibitem{Komatsu:2008hk}
Komatsu E {\em et~al.\/} (WMAP Collaboration) 2009 {\em Astrophys.J.Suppl.\/}
  {\bf 180} 330--376 (\textit{Preprint} \eprint{arXiv:0803.0547})

\bibitem{1994CQGra..11..443G}
{Gourgoulhon} E and {Bonazzola} S 1994 {\em Classical and Quantum Gravity\/}
  {\bf 11} 443--452

\bibitem{Kolb:2005me}
{Kolb} E~W, {Matarrese} S, {Notari} A and {Riotto} A 2005 {\em ArXiv
  e-prints\/} (\textit{Preprint} \eprint{arXiv:hep-th/0503117})

\bibitem{Crocce:2005xy}
Crocce M and Scoccimarro R 2006 {\em Phys.Rev.\/} {\bf D73} 063519
  (\textit{Preprint} \eprint{astro-ph/0509418})

\bibitem{Crocce:2007dt}
Crocce M and Scoccimarro R 2008 {\em Phys.Rev.\/} {\bf D77} 023533
  (\textit{Preprint} \eprint{0704.2783})

\bibitem{Ellis:1998ct}
Ellis G~F~R and van Elst H 1999 {\em NATO Adv. Study Inst. Ser. C. Math. Phys.
  Sci.\/} {\bf 541} 1--116 (\textit{Preprint} \eprint{gr-qc/9812046})

\bibitem{Tsagas:2007yx}
Tsagas C~G, Challinor A and Maartens R 2008 {\em Phys.Rept.\/} {\bf 465}
  61--147 * Brief entry * (\textit{Preprint} \eprint{0705.4397})

\bibitem{2000AnPhy.282..285G}
{Gebbie} T and {Ellis} G~F~R 2000 {\em Annals of Physics\/} {\bf 282} 285--320
  (\textit{Preprint} \eprint{arXiv:astro-ph/9804316})

\bibitem{Pirani}
Pirani F~A~E 1964 {\em Introduction to gravitational radiation Theory. Lectures
  in GR, Brandeis Lectures\/} Master's thesis Ed. Deser and Ford,

\bibitem{1981MNRAS.194..439T}
{Thorne} K~S 1981 {\em \mnras\/} {\bf 194} 439--473

\bibitem{Thorne:1980ru}
Thorne K 1980 {\em Rev.Mod.Phys.\/} {\bf 52} 299--339

\bibitem{1983AnPhy.150..455E}
{Ellis} G~F~R, {Matravers} D~R and {Treciokas} R 1983 {\em Annals of Physics\/}
  {\bf 150} 455--486

\end{thebibliography}

\providecommand{\newblock}{}

\end{document}